\documentclass[aps,preprintnumbers,11pt,amsmath,amssymb,nofootinbib]{revtex4}

\usepackage{braket}
\usepackage{float}
\usepackage{graphics,epsfig}
\usepackage{epstopdf}
\usepackage{bm}
\usepackage{slashed}
\usepackage{graphicx}
\usepackage{amsmath}
\usepackage{amsfonts}
\usepackage{amssymb}
\usepackage{color}%
\usepackage{ulem}
\usepackage{dcolumn}
\usepackage[
            pdfstartview=FitH,
            bookmarksnumbered=true,
            bookmarksopen=true,
            colorlinks,
            linkcolor=blue,
            anchorcolor=green,
            citecolor=red
            ]{hyperref}
\usepackage{enumerate}
\providecommand{\U}[1]{\protect\rule{.1in}{.1in}}

\begin{document}

 \newcommand{\bq}{\begin{equation}}
 \newcommand{\eq}{\end{equation}}
 \newcommand{\bqn}{\begin{eqnarray}}
 \newcommand{\eqn}{\end{eqnarray}}
 \newcommand{\nb}{\nonumber}
 \newcommand{\lb}{\label}
 \newcommand{\calp}{{\cal{P}}_{+}}
 \newcommand{\calpt}{{\cal{P}}_{\times}}
 \newcommand{\calps}{{\cal{P}}_{\sigma}}
 \newcommand{\calr}{{\cal{R}}}
 \newcommand{\calpp}{{\cal{P}}_{+,v}}
 \newcommand{\calpsv}{{\cal{P}}_{\sigma,v}}
 \newcommand{\calptp}{{\cal{P}}_{\times,v}}
 \newcommand{\calpsvv}{{\cal{P}}_{{\sigma},vv}}
 \newcommand{\calppp}{{\cal{P}}_{+,vv}}
 \newcommand{\calptpp}{{\cal{P}}_{\times,vv}}
 \newcommand{\calh}{{\cal{H}}}
 \newcommand{\calhp}{{\cal{H} '}}
 \newcommand{\calt}{{\cal{T}}_{+}}
 \newcommand{\caltt}{{\cal{T}}_{\times}}
 \newcommand{\calts}{{\cal{T}}_{\sigma}}
 \newcommand{\caltu}{{\cal{T}}_{+,u}}
 \newcommand{\calttu}{{\cal{T}}_{\times,u}}
 \newcommand{\caltsu}{{\cal{T}}_{\sigma,u}}
 \newcommand{\caltuu}{{\cal{T}}_{+,uu}}
 \newcommand{\calttuu}{{\cal{T}}_{\times,uu}}
 \newcommand{\caltsuu}{{\cal{T}}_{\sigma,uu}}
 \newcommand{\caltv}{{\cal{T}}_{+,v}}
 \newcommand{\calttv}{{\cal{T}}_{\times,v}}
 \newcommand{\caltsv}{{\cal{T}}_{\sigma,v}}
 \newcommand{\caltvv}{{\cal{T}}_{+,vv}}
 \newcommand{\calttvv}{{\cal{T}}_{\times,vv}}
 \newcommand{\caltsvv}{{\cal{T}}_{\sigma,vv}}
\newcommand{\PRL}{Phys. Rev. Lett.}
\newcommand{\PL}{Phys. Lett.}
\newcommand{\PR}{Phys. Rev.}
\newcommand{\PRD}{Phys. Rev. D.}
\newcommand{\CQG}{Class. Quantum Grav.}
\newcommand{\JCAP}{J. Cosmol. Astropart. Phys.}
\newcommand{\JHEP}{J. High. Energy. Phys.}

\baselineskip=0.6 cm \title{Rotations of the polarization of a gravitational wave propagating  in Universe}
\author{Jia-Xi Feng $^{1,2}$}
\thanks{fengjiaxigw@gmail.com}
\author{Fu-Wen Shu$^{1,2}$}
\thanks{shufuwen@ncu.edu.cn; the corresponding author}
\author{Anzhong Wang$^{3,4}$}
\thanks{Anzhong$\_$Wang@baylor.edu}

\affiliation{
$^{1}$Department of Physics, Nanchang University, Nanchang, 330031, China\\
$^{2}$Center for Relativistic Astrophysics and High Energy Physics, Nanchang University, Nanchang 330031, China\\
$^{3}$Institute for Advanced Physics $\&$ Mathematics, Zhejiang University of Technology, Hangzhou, 310032, China\\
$^{4}$GCAP-CASPER, Physics Department, Baylor University, Waco, TX 76798-7316, USA}

\vspace*{0.2cm}
\begin{abstract}

In this paper, we study the polarization of a gravitational wave (GW) emitted by an astrophysical source at a cosmic distance propagating
 through the Friedmann-Lema\^itre-Robertson-Walk (FLRW) universe. By considering the null geodesic deviations, we first provide a definition
 of the polarization of the GW in terms of the Weyl scalars with respect to a parallel-transported frame along the null geodesics, and then
 show explicitly that,  due to different effects of the expansion of the universe  on the two polarization modes, the so-called ``+" and ``$\times$"
 modes,  the polarization angle of the GW changes generically, when it is propagating through the curved background.  More precisely,
 due to the presence of the matter field of the FLRW background, both of the ``$\times$'' and ``+'' modes
can get amplified or diluted, depending on their waveforms, so in principle the effects to
the ``$\times$'' modes are different from those to the ``+'' modes. As a result, the polarization angle will change along the wave path, regardless of which source it is. As a concrete example, we directly compute the polarization angle in a binary system, and show that different epochs,  radiation-, matter- and $\Lambda$-dominated,  have different effects on the polarization.
 In particular, for a GW emitted by a binary system, we find explicitly the relation between the change of the polarization angle $|\Delta \varphi|$
 and the redshift  $z_s$ of the source in different epochs. In the  $\Lambda$CDM model, we find that the order of $|\Delta \varphi|{\eta_0 F}$
  is typically $O(10^{-3})$ to $O(10^3)$, depending on the values of $z_s$, where $\eta_0$ is the (comoving)  time of the current universe, and
  $F\equiv\Big(\frac{5}{256}\frac{1}{\tau_{obs}}\Big)^{3/8}\left(G_NM_c\right)^{-5/8}$ with $\tau_{obs}$ and  $M_c$ being, respectively, the time
  to coalescence in the observer's frame and the chirp mass of the binary system.  The typical value of $|\Delta\varphi|$ for LIGO-Virgo sources is $10^{-21}$. Hence,  it may not be easily detected with current detectors.

\end{abstract}
\maketitle
\newpage
\vspace*{0.2cm}

\maketitle

\section{Introduction }\label{1}
\renewcommand{\theequation}{1.\arabic{equation}} \setcounter{equation}{0}

With the detection of gravitational waves (GWs) by the LIGO  collaboration  \cite{GW150914,GW151226}, a new era, the gravitational wave astronomy,
began, after exactly 100 years since Einstein first predicted the existence of GWs \cite{Einstein1916}  by using his brand new theory of general relativity, established
only one year earlier. The masses of the two binary black holes (BBHs) in the event GW151226 were,
respectively, $14 M_{\odot}$ and $7 M_{\odot}$  \cite{GW151226}, well in the range of stellar-mass black holes known so far \cite{BHs}. On the other hand, in the
event GW150914  \cite{GW150914},  the masses of the two black holes were, respectively, $36 M_{\odot}$ and $29 M_{\odot}$,
which are much more massive than the known ones \cite{BHs}. This has already stimulated lots of interest and various scenarios have been proposed, including
their formation from very massive stars ($50 - 100 M_{\odot}$) \cite{MassiveStars},  PopIII stars \cite{PopIIIStars},  and primordial black holes \cite{PBHs}(see also\cite{T.Z. Wang}).
Despite
of the difference among the masses of BBHs in these two events, the distances of them to  Earth are almost the same, about $400$Mpc, which corresponds to a
redshift of $z \simeq 0.09$ \cite{GW150914,GW151226}.
After that, LIGO/Virgo scientific collaborations detected dozen more GWs \cite{GW190814,2019,GWs19a,GWs19b,Abbott:2017vtc,Abbott:2017gyy,Abbott:2017oio,TheLIGOScientific:2017qsa},
 including possibly the coalescence of  neutron-star (NS)/BH binary, although some  details of these detections have not been  released yet \cite{LIGO}.
Over the next few years, the advanced LIGO and Virgo will continuously increase their sensitivities, and
once at their designed goals, together with other detectors of
the second generation, such as KAGRA \cite{KAGRA}, they could be able to detect heavy BBHs up to redshifts of unity.  The third generation detectors, both ground and space based, such as
the Einstein Telescope \cite{ET}, Cosmic Explorer \cite{CE},  LISA \cite{LISA}, Taiji \cite{Taiji}, Tianqin \cite{Tianqin}, and DECIGO \cite{DEIGO},
will substantially increase their sensitivities and allow us to detect GWs from sources up to redshifts
of $z \simeq 20$, well within the epoch of reionization \cite{VE}.  Therefore, the studies of such GWs and their sources, among other things,  will open a new window
to explore the early universe.

Such studies
are normally divided into two different zones, the generation and propagation zones \cite{YYP}. The propagation zone is further divided into three different phases,
the inspiral, plunge/merger and ringdown phases.  In the inspiral phase, the post-Newtonian (PN) approximations are usually used, while in merger phase,
the gravitational field is very strong and the field equations (of Einstein as well as of other theories), become highly nonlinear, and  heavy numerical
calculations  are often inevitably used. In the ringdown phase, quasi-normal modes (QNM) analysis is found sufficient.

When GWs are far from their sources
(in the far propagation zone), they become very weak, and can be considered as linear perturbations. Previous studies were mainly focused on perturbations in the
Minkowski background \cite{GWs}, but recently studies of perturbations in the Friedmann-Lema\^itre-Robertson-Walk (FLRW) background have started to attract
attention \cite{ABK15,Chu15,BGY15,KR16,TW16} (See also \cite{DBBE}).


In particular,
for a GW generated by an astrophysical source with a distribution of matter in a finite region, such as an inspiraling compact binary system, it can be considered as
perturbations on the FLRW background $\hat{g}_{\mu\nu} \equiv a^2\eta_{\mu\nu}$\footnote{In this paper, we shall use quantities with hats to denote the
background ones, and reserve the ones with bars for complex conjugates.},
\bq
\lb{eq1.1}
g_{\mu\nu} = a^2(\eta)\left(\eta_{\mu\nu} + \epsilon h_{\mu\nu}\right),
\eq
where $\eta_{\mu\nu} = {\mbox{diag.}}\left(1, -1, -1, -1\right)$,  $\eta$ denotes the conformal time of the FLRW universe, and
$\left|\epsilon h_{\mu\nu}\right| \ll 1$.
In this paper, we shall use the notations and conventions of \cite{dInverno}, and in particular, we shall set the speed of light to unity, $c=1$. In the transverse-traceless (TT) gauge, 
\bq
\lb{eq1.2}
h_{\mu\nu}dx^{\mu}dx^{\nu} = D_{ij}dx^idx^j,
\eq
with $\delta^{ij}D_{ij} = 0 = \partial^{i}D_{ij}$, the linear perturbations  $D_{ij}$'s equations are  given by \cite{Chu15},
\bq
\lb{eq1.3}
{\Box} D_{ij} = 16\pi G_N \Pi_{ij},
\eq
where ${\Box} \equiv \hat{g}^{\mu\nu}\hat{\nabla}_{\mu}\hat{\nabla}_{\nu}$, $\hat{\nabla}_{\mu}$ denotes the covariant derivative with respect to $\hat{g}_{\mu\nu}$,
and $ \Pi_{ij}$ denotes the part of matter perturbations $\delta T^i_j$, subjected to the TT constraints,
$\delta^{ij}\Pi_{ij} = 0 = \partial^{i}\Pi_{ij}$. When far from the source, $ \left|\vec{x}\right| \gg D$, where $D$ denotes the spatial extent
of the source,
 and $\left|\vec{x}\right| \; \left[\equiv \sqrt{x^2 + y^2 + z^2}\right]$ is the comoving distance between the center of the source and the observer.

\begin{figure}
\centering
\includegraphics[scale=0.45]{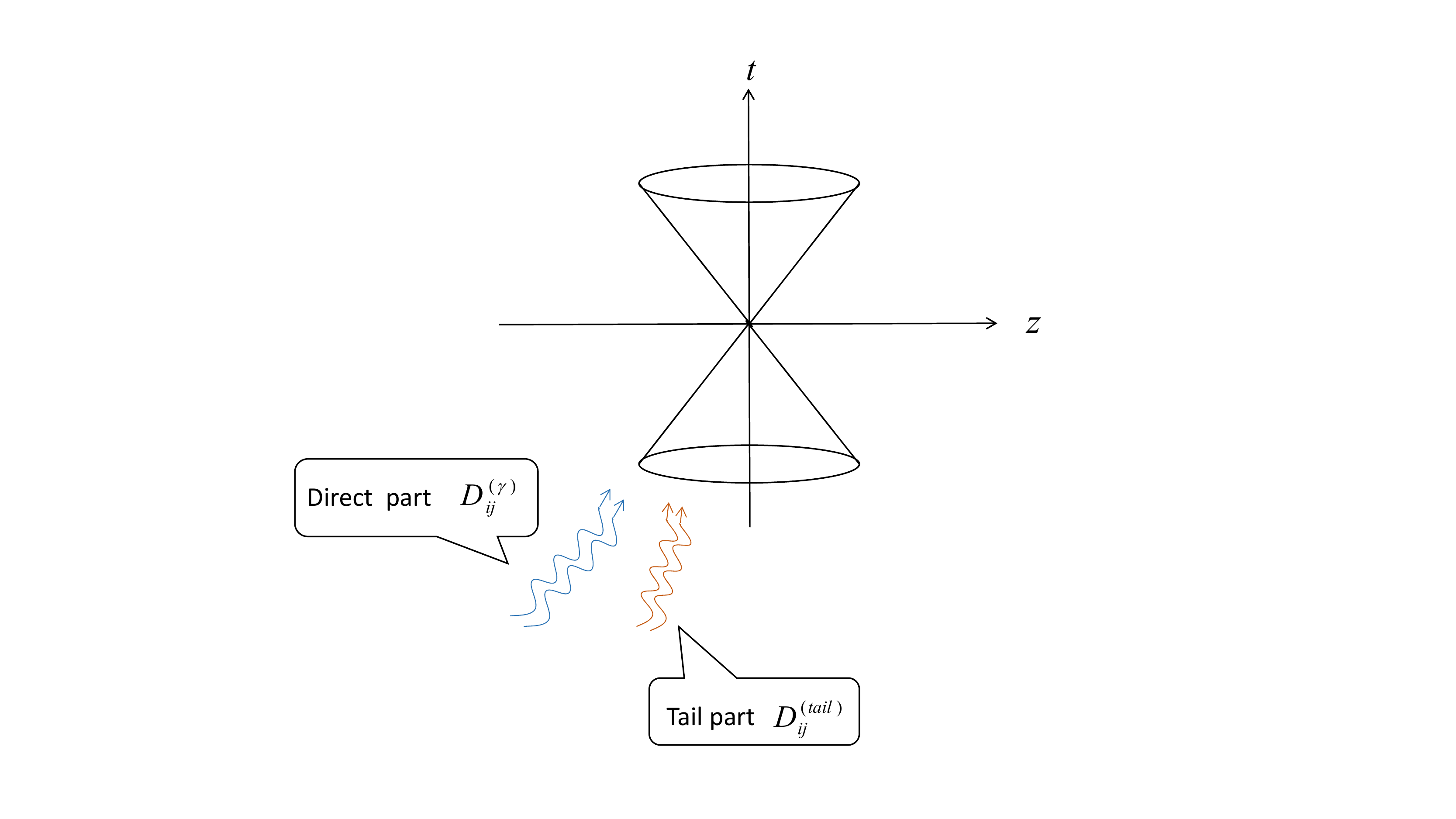}
\caption{$D_{ij}$ consists of two parts: one is the propagation (or direct) part $D_{ij}^{(\gamma)}$ which travels along the light-cone,
 and the other is the tail part $D^{(tail)}_{ij}$ which travels inside the light-cone.} \label{fig:Ptz}
\end{figure}

In the FLRW background \eqref{eq1.1}, when GWs propagate through this curved  space-time, an additional term appeared due to the scattering off the curved backgorund,
the ``backscattering waves'' or ``tails'' \footnote{In what follows we call this scattering the backscattering of the GWs over the FLRW background. \cite{Chu15,Chu:2011ip}. Namely,} $D_{ij}$ consists of two parts: one
 travels along the light-cone, denoted by $D_{ij}^{(\gamma)}$, which will be referred to as the propagation (or direct) part, and the other travels inside the light-cone,
 denoted by $D^{(tail)}_{ij}$, the so-called tails as shown in Fig. \ref{fig:Ptz}. They are given, respectively, by,
\bq
\lb{eq1.4}
D_{ij}^{(\gamma)} =  \frac{ 4G_N F_{ij}^{(\gamma)}(v)}{ \left|\vec{x}\right| a(\eta)}, \;\;\;
D_{ij}^{(tail)} = \frac{4G_N F_{ij}^{(tail)}(\eta, \vec{x})}{a(\eta)},
\eq
where $v \equiv \frac1{\sqrt{2}}(\eta - \left|\vec{x}\right|)$, and
\bqn
\lb{eq1.5}
F_{ij}^{(\gamma)} (v) &\equiv& \int_{\mathbb{R}^3}{\mathrm{d}^3\vec{x}'a^3(v) \Pi_{ij}\left(v, \vec{x}'\right)},\nb\\
F_{ij}^{(tail)}(\eta, \vec{x}) &\equiv&  \int{a^3(\eta')\mathrm{d}\eta'}
 \int_{\mathbb{R}^3}{\mathrm{d}^3\vec{x}' \Theta(\hat\sigma) \Pi_{ij}\left(\eta', \vec{x}'\right)} \times \frac{\partial J(\eta, \eta'; \hat\sigma)}{\partial \hat\sigma},
\eqn
where $\mathbb{R}^3$ is the region of source, and $ \Theta(x)$ denotes  the Heaviside function,
\bq
\lb{eq1.5a}
\Theta(x) = \begin{cases}
1, & x \ge 0,\cr
0, & x < 0.\cr
\end{cases}
\eq
And  $\hat\sigma \equiv \left[(\eta - \eta')^2 - (z-z')^2 -  (\vec{x}_{\bot} - \vec{x}_{\bot}')^2\right]/2, \; \vec{x} \equiv (z, \vec{x}_{\bot}),\;  \vec{x}_{\bot} \equiv (x, y)$.
The function $J(\eta, \eta'; \hat\sigma)$ satisfies the equation,
\bq
\lb{eq1.6}
{\cal{W}}_2{(\eta, {z})} J(\eta, \eta'; \hat\sigma) = {\cal{W}}_2{(\eta', {z}')} J(\eta, \eta'; \hat\sigma) = 0,
\eq
with
\bq
\lb{eq1.7}
{\cal{W}}_2{(\eta, z)} \equiv \partial^2_{\eta} - \partial^2_{z} - \frac{a''(\eta)}{a(\eta)},
\eq
where $a''(\eta) \equiv d^2a(\eta)/d\eta^2$, etc. When $a(\eta)$ is a constant, for which the background becomes the Minkowski spacetime,
the last term of Eq.(\ref{eq1.7}) vanishes, so Eq.(\ref{eq1.6}) reduces to $(\partial^2_{\eta} - \partial^2_{z})J(\eta, \eta'; \hat\sigma)= 0$, which represents
waves propagating along the light-cone $\eta = \pm z$, that is, the tail part   reduces to the direct part, and the corresponding GWs all move exactly along the light-cones.

 In this paper, we shall study the polarizations of GWs produced by an astrophysical source, which propagate
 through our universe over a cosmic distance. As a first step, we shall assume that our universe
is homogeneous and isotropic. The efforts of the inhomogeneity will be considered somewhere else \cite{JF20}.

The rest of this paper is organized as follows. In Section II, assuming that the GWs are far from their sources, we show how
to define  the polarization angle of the  GW  propagating through  the flat FLRW background at a cosmic distance. In Section III, different epochs, in which the universe is dominated by different components
of matter field, are considered. In particular, we calculate explicitly the changes of  the polarization angle in each of these epochs.  Our main conclusions are summarized in section IV.

There are also four appendices, \ref{appda}, \ref{appdb1}, \ref{appdb} and \ref{appdc}, in which some mathematical calculations involved  in this paper are given in detail. In particular, in Appendix \ref{appdb} we consider the timelike geodesics deviations,
from which we define the polarization of a GW propagating through the flat FLRW universe, and show that it is  the same as that obtained by considering null geodesic deviations studied
in Section II.

\section{Polarizations of Gravitational Waves}\label{2}
\renewcommand{\theequation}{2.\arabic{equation}} \setcounter{equation}{0}

In the rest of this paper, we shall assume that the GWs are far from their sources, so they are well-described by Eqs.(\ref{eq1.4})-(\ref{eq1.7}).
In addition, we shall choose the coordinates so that the $z$-axis is passing through  the observer and  the center of the source. Then,
the transverse-traceless conditions lead to,
\bqn
\lb{2.1}
D_{22} &=& -D_{33} \equiv h_{+}, \quad   D_{23} = D_{32} \equiv - h_{\times},   \nb\\
D_{i1} &=& 0,\; (i = 1, 2, 3),
\eqn
with
\bqn
\lb{2.2}
h_\sigma(\eta, \vec{x}) \equiv \frac{ {\cal{P}}_\sigma(v)}{{\cal{R}}}  + \frac{{\cal{T}}_\sigma(u,v)}{a^n} + {\cal{O}}\left(\frac{1}{|\vec{x}|^2}\right),
\eqn
where $\sigma = +, \times, \; u \equiv {\frac1{\sqrt{2}}}(\eta + \left|\vec{x}\right|)$, and ${\cal{R}} \equiv a(\eta) |\vec{x}|$ is the physical distance from the center of the source to the observer,
  $n$ is a constant, and
 \bqn
\lb{eq2.2a}
 {\cal{T}}_\sigma(u,v) \propto
 \int_{\mathbb{R}^3}{\mathrm{d}^3\vec{x}' \Theta(\hat\sigma) \Pi_{ij}\left(\eta', \vec{x}'\right)}
  \frac{\partial J(\eta, \eta'; \hat\sigma)}{\partial \hat\sigma}. ~~~
\eqn
In the radiation-dominated epoch we have ${\cal{T}}_\sigma(u, v) = 0$, while in the de Sitter spacetime,
or  the matter-dominated  epoch, we have ${\cal{T}}_\sigma(u, v) \not= 0$,  but with $n = 0, 3/2$, respectively \cite{Chu15}.

 In writing the above expressions, we expanded the direct part $D_{ij}^{(\gamma)}(t, \vec{x})$ only to the first-order of $1/|\vec{x}|$.
$ {\cal{P}}_\sigma(v)$ is related to the integration of the direct part of the source via the relation,
\bq
\lb{2.3}
 {\cal{P}}_\sigma(v) \simeq   G_N \int_{\mathbb{R}^3}{\mathrm{d}^3\vec{x}'a^3(v) \Pi_{ij}\left(v, \vec{x}'\right)},
 \eq
 which propagates along the light-cone, $v=$ Constant,  toward  the observer in the increasing direction of $z$. On the other hand,  $ {\cal{T}}_\sigma(u,v)$
 denotes  the tails propagating inside the light-cone, due to the backscattering of the GWs over the FLRW background.  Note that we keep this part to the linear order
 of ${\cal{T}}_\sigma$ without expanding it further in terms of $1/|\vec{x}|$.

Considering Eqs.(\ref{eq1.1}), (\ref{eq1.2}) and (\ref{2.1}), we find that the metric of such space-times takes the form,
\bqn\label{gwmetric}
\mathrm{d}s^2&=&e^{-M}\left(\mathrm{d}\eta^2 - \mathrm{d}z^2\right) -e^{-U}\left\{e^{V}\cosh{W}\mathrm{d}x^2-2\sinh{W}\mathrm{d}x\mathrm{d}y +e^{-V}\cosh{W}\mathrm{d}y^2\right\},
\eqn
where
\begin{eqnarray}
M &=& U = - 2\ln a, \quad V= \epsilon h_+, \quad W=\epsilon h_{\times}.
\end{eqnarray}
 Then, the analysis of propagation and polarizations of GWs in the Universe can follow the one given in \cite{Szekeres72,Griffiths76a,Griffiths:1991zp,WThesis,Wang91,HFZ19}
 \footnote{It should be noted that in \cite{Szekeres72,Griffiths76a,Griffiths:1991zp,WThesis,Wang91} only plane gravitational waves were considered. In this paper,
 as to be shown below, such studies can be generalized to the current case, in which the gravitational waves are no longer plane waves, due to the  curved background, as can be
 seen explicitly from Eqs.(\ref{eq1.4}) - (\ref{eq1.7}).}, which is crucially based on the following decompositions of the Riemann tensor,
\begin{eqnarray}
R_{\mu\nu\lambda\beta} &=& C_{\mu\nu\lambda\beta} + \frac{R}{6}\left(g_{\mu\beta}g_{\nu\lambda} - g_{\mu\lambda}g_{\nu\beta}\right) + \frac{1}{2}\Big(g_{\mu\lambda}R_{\nu\beta}
+ g_{\nu\beta}R_{\mu\lambda}
- g_{\nu\lambda}R_{\mu\beta} - g_{\mu\beta} R_{\nu\lambda}\Big),
\end{eqnarray}
where $R_{\mu\nu}$ and $R$ denote, respectively, the Ricci tensor and scalar, and $C_{\mu\nu\lambda\beta}$ is the Weyl tensor. The latter is considered as
representing pure gravitational fields, while $R_{\mu\nu}$ is connected to the energy-momentum tensor $T_{\mu\nu}$ through the
Einstein field equations $R_{\mu\nu} = 8\pi G_N\left( T_{\mu\nu} - g_{\mu\nu}T/2\right)$, which is usually considered as the spacetime curvature produced by matter.

In terms of $C_{\mu\nu\lambda\beta}$ and $R_{\mu\nu}$, the Bianchi identities take the form,
\begin{eqnarray}
\nabla^{\alpha}C_{\mu\nu\lambda\alpha} = J_{\mu\nu\lambda},
\end{eqnarray}
which represents the interaction between matter and pure gravitational  fields (represented by the Weyl tensor $C_{\mu\nu\lambda\beta}$), where
\begin{eqnarray}
 J_{\mu\nu\lambda} \equiv \nabla_{[\nu}R_{\mu]\lambda} - \frac{1}{6}g_{\lambda[\mu}\nabla_{\nu]}R,
 \end{eqnarray}
is directly related to the energy-momentum tensor $T_{\mu\nu}$ through the Einstein field equations, with $[A, B] \equiv (AB-BA)/2$. Clearly, in vacuum we have
$J_{\mu\nu\lambda} = 0$.

To study the phenomena, it is found very convenient to use the Newman-Penrose formula \cite{NP62},  a tetrad formula but with a particular choice of the tetrad,
\begin{eqnarray}
l^{\mu} &\equiv& B\left(\delta^{\mu}_{\eta} - \delta^{\mu}_{z}\right), \quad n^{\mu} \equiv A \left(\delta^{\mu}_{\eta} + \delta^{\mu}_{z}\right),\nb\\
m^{\mu} &\equiv& \zeta^2 \delta^{\mu}_{2} + \zeta^3 \delta^{\mu}_{3},  \quad
\bar{m}^{\mu} \equiv \overline{m^{\mu}},
\end{eqnarray}
where an overline/bar  denotes the complex conjugate, and
\begin{eqnarray}
 \zeta^2 &\equiv& \frac{1  {-} {\cal{G}}}{\sqrt{2} a}, \;\;
  \zeta^3 \equiv \frac{i(1  {+} {\cal{G}})}{\sqrt{2} a},   \nb\\
  AB &\equiv& \frac{1}{2a^2}, \quad
  {\cal{G}} \equiv \frac{\epsilon}{2}\left(h_{+} - i h_{\times}\right).
\end{eqnarray}

It can be shown that, to the leading order of $\epsilon$,  each of  $l^{\mu}$ and $n^{\mu}$ defines a null geodesic congruence,
\begin{eqnarray}
l^{\nu}\nabla_{\nu} l^{\mu} &=& - \frac{B A_{,\eta}}{A} l^{\mu} + {\cal{O}}\left(\epsilon^2\right), \nb\\
n^{\nu}\nabla_{\nu} n^{\mu} &=& - \frac{A B_{,\eta}}{B} n^{\mu} + {\cal{O}}\left(\epsilon^2\right),
\end{eqnarray}
where $\nabla_{\nu}$ denotes the covariant derivative with respect to $g_{\mu\nu}$, and $A_{,\eta} \equiv \partial A/\partial \eta$, etc.
Projecting the Weyl and Ricci tensors onto the above null tetrad, we find that the ten independent components of the Weyl tensor  are given by
the five Weyl (complex) scalars, $\Psi_A$, while the ten  independent components of the Ricci tenor are given by the seven Ricci scalar,
$\Phi_{AB}$, given explicitly in Appendix \ref{appda} for the metric Eq.(\ref{gwmetric}). In particular, we have
\bqn
\lb{RePsi0}
{\mbox{Re}}\left(\Psi_0\right) &=& {-}\frac{1}{8 A^2 a^4}\Bigg\{\frac{1}{\calr}\Big[2{\calppp}-2\sqrt{2}{\calh} {\calpp} +({\calh}^{2}-{\calhp}){\calp}\Big]\nb\\
&&~~~ +\frac{1}{a^n}\Bigg[2{\caltvv}\left(1-\frac{x^2+y^2}{2z^2}\right) +\sqrt{2}n{\calh}\Bigg(({\caltu}-{\caltv}) \left(1-\frac{x^2+y^2}{2z^2}\right)\nb\\
&&~~~-({\caltu}+{\caltv})\Bigg)+n(n{\calh}^2-{\calhp}){\calt}\Bigg]\Bigg\}\epsilon+ {\cal{O}}\left(\epsilon^2\right),
\eqn
\bqn
\lb{ImPsi0}
{\mbox{Im}}\left(\Psi_0\right) &=& \frac{1}{8 A^2 a^4}\Bigg\{\frac{1}{\calr}\Bigg[2{\calptpp}-2\sqrt{2}{\calh} {\calptp} +({\calh}^{2}-{\calhp}){\calpt}\Bigg] \nb\\
&&~+\frac{1}{a^n}\Bigg[2{\calttvv}\left(1-\frac{x^2+y^2}{2z^2}\right)+\sqrt{2}n{\calh}\Bigg(({\calttu}-{\calttv})\left(1-\frac{x^2+y^2}{2z^2}\right)\nb\\
&&~-({\calttu}+{\calttv})\Bigg)+n(n{\calh}^2-{\calhp}){\caltt}\Bigg]\Bigg\}\epsilon + {\cal{O}}\left(\epsilon^2\right),
\eqn
where
\bq
\lb{eq2.16}
\calpsv=\frac{\mathrm{d}\calps(v)}{\mathrm{d}v}, \;\;\;\;
\calh=\frac{a '(\eta)}{a(\eta)},\;\;\;\;
\calh '=\frac{\mathrm{d}\calh}{\mathrm{d}\eta}.
\eq
%

The importance of the above decompositions is that each component of these scalars has its own physical interpretation. For example,  $\Psi_0$ and  $\Psi_4$ represent
the gravitational waves propagating along the null geodesics, defined by $l^{\mu}$ and $n^{\mu}$, respectively, while  $\Psi_1$ and  $\Psi_3$ represent the longitudinal components,
and $\Psi_2$ the Coulomb  component  \cite{Szekeres65}.

\subsection{Polarizations of GWs}

To study the polarization of GWs propagating in the FRLW background, let us consider null geodesic deviations. The main motivation of studying null geodesic deviations, instead of those timelike which are related to observations directly,  is that GWs are traveling along their null geodesics before catching by the detectors. As shown in \cite{Wang91,WThesis}, the considerations of   null geodesic congruences will lead to the same definition for the polarization of a given GW. However, in  \cite{Wang91,WThesis} only   plane gravitational wave (Petrov Type N)spacetimes were considered, and in what follows we will show that this is also true for the spacetimes considered here, by considering null geodesic deviations. In fact,  in Appendix \ref{appdb} we show that the definition of the polarization of a GW should be valid in any given spacetime, independent of its symmetry and the nature of the geodesic deviations, null or timelike.

To our purpose,  in the following let us  consider the null geodesic congruences formed by $l^{\mu}$.  We construct the following four unity  vectors,
 \bqn
 \lb{B.1}
e^{\mu}_0 &\equiv&t^{\mu} \equiv\frac1{\sqrt{2}}(n^{\mu}+l^{\mu}), \quad
e_1^{\mu} \equiv s^{\mu} \equiv \frac1{\sqrt{2}}(n^{\mu}-l^{\mu}),\nb\\
e^{\mu}_2 &\equiv& \frac{1}{\sqrt{2}}\left(m^{\mu} + \bar{m}^{\mu}\right),\quad
e^{\mu}_3 \equiv - \frac{i}{\sqrt{2}}\left(m^{\mu} - \bar{m}^{\mu}\right),\nb\\
\eqn
so that they form an orthogonal base
\bq
\lb{B.2}
g_{\mu\nu} e^{\mu}_a e^{\nu}_b = \eta_{ab},
\eq
where $e^{\mu}_a \equiv \left(t^{\mu}, s^{\mu}, e^{\mu}_2, e^{\mu}_3\right), (a = 0, 1, 2, 3)$,
and $\eta_{ab} = \text{diag.}\left(1, -1, -1, -1\right)$.

 Let $\eta^{\mu}$ be the geodesic deviation between two neighbor geodesics and $\eta^{\mu} l_{\mu} = 0$. Then, after some tedious but straightforward calculations, we find that the null geodesic deviation is given by,
\bqn
\lb{geodesic1}
\frac{\mathrm{d}^2\eta^{\mu}}{\mathrm{d}\lambda^2} &=& - R^{\mu}_{\nu\lambda\beta} l^{\nu}\eta^{\lambda}l^{\beta} \nb\\
& =&\Big\{\Big[\left(\Psi_2+\bar{\Psi}_2\right)-2\Phi_{11}+2\Lambda\Big]{(e^{\mu}_0-e^{\nu}_1)(e^{\nu}_0-e^{\mu}_1)}\nb\\
&&{-} \Phi_{00} e^{\mu\nu}_{o} +{\mbox{Re}} \Psi_0\; e^{\mu\nu}_{+}  +  {\mbox{Im}} \Psi_0\; e^{\mu\nu}_{\times}\nb\\
&& {+} \sqrt{2}\Big[{\mbox{Re}}\left(\Phi_{01} {-}  \Psi_1\right) e^{\mu\nu}_{l,2}  +{\mbox{Im}}  \left(\Phi_{01}  {-}   \Psi_1\right)e^{\mu \nu}_{l,3}\Big]\Big\} \eta_{\nu}, \nb\\
\eqn
 where
\bqn
e^{\mu\nu}_{+} &\equiv&  e^{\mu}_2e^{\nu}_2 -  e^{\mu}_3e^{\nu}_3,\quad
e^{\mu\nu}_{\times} \equiv  e^{\mu}_2e^{\nu}_3 +  e^{\mu}_3e^{\nu}_2,\nb\\
e^{\mu\nu}_{l,2} &\equiv&e^{\mu}_2 l^{\nu} + e^{\nu}_2 l^{\mu},\quad
e^{\mu\nu}_{l,3}\equiv e^{\mu}_{3}l^{\nu} + e^{\nu}_{3}l^{\mu},\quad
e^{\mu\nu}_{o} \equiv e^{\mu}_2e^{\nu}_2 +   e^{\mu}_3e^{\nu}_3.
\eqn
The first term of Eq.\eqref{geodesic1} appears due to the  background that is curved,  in which $h_{\sigma}$ is the function of $u$  and $ v$. If the background is flat, then $h_{\sigma}$ is a function of $u$ or $ v$ only, and, as a result,  this term will vanish.
For matter that satisfies the energy conditions \cite{HE73}, we always have
$\Phi_{00} > 0$. Thus, the second term   always makes the geodesic congruence uniformly contracting in the plane orthogonal to the propagation of the GWs,
spanned by $e_2^{\mu}$ and $e_3^{\mu}$.

In contrast to matter, the GW has different effects. In particular, the ${\mbox{Re}}\left(\Psi_0\right)$ part
stretches the $e_2$-direction and meantime squeezes the $e_3$-direction or vice versa, depending on the sign of
${\mbox{Re}}\left(\Psi_0\right)$.  The ${\mbox{Im}}\left(\Psi_0\right)$ part does the same, but now along the axes
$e_2'$ and $e_3'$, which are obtained by a  $45^0$ rotation in the ($e_2, e_3$)-plane.
On the other hand, making the following rotation,
\bqn
\lb{rotation1}
e_2^{\mu}  &=& \cos\varphi_0 \hat{e}_2^{\mu} + \sin\varphi_0 \hat{e}_3^{\mu},\nb\\
e_{3}^{\mu}  &=& - \sin\varphi_0 \hat{e}_2^{\mu} + \cos\varphi_0 \hat{e}_3^{\mu},
\eqn
where
\bq
\lb{varphi0}
\tan 2\varphi_0 \equiv - \frac{{\mbox{Im}}\left(\Psi_0\right)}{{\mbox{Re}}\left(\Psi_0\right)}.
\eq	
Eq.(\ref{geodesic1}) becomes,
\bqn
\lb{geodesic1.1}
&& \frac{\mathrm{d}^2\eta^{\mu}}{\mathrm{d}\lambda^2}
 = \Big\{\Big[(\Psi_2+\bar{\Psi}_2)-2\Phi_{11}+2\Lambda\Big]{(e^{\mu}_0-e^{\nu}_1)(e^{\nu}_0-e^{\mu}_1)}\nb\\
&& ~~~~~~~~~~~ {-} \Phi_{00} \hat e^{\mu\nu}_{o} +\Big(\Psi_0\bar{\Psi}_0\Big)^{1/2} \hat{e}^{\mu\nu}_{+}\nb\\
&& ~~~~~~~~~~~{+}  \sqrt{2}\Big[{\mbox{Re}}\left(\Phi_{01} {-}  \Psi_1\right)(\sin\varphi_0\hat e^{\mu\nu}_{l,3}+\cos\varphi_0\hat e^{\mu \nu}_{l,2})\nb\\
&&~~~~~~~~~~~ +{\mbox{Im}}  \left(\Phi_{01}  {-}   \Psi_1\right)(\cos\varphi_0\hat e^{\mu\nu}_{l,3}-\sin\varphi_0\hat e^{\mu \nu}_{l,2})\Big]\Big\} \eta_{\nu},
\eqn
where
\bqn
e^{\mu\nu}_{+}&=&\cos 2\varphi_0 \hat{e}^{\mu\nu}_{+}+\sin 2\varphi_0\hat{e}^{\mu\nu}_{\times},\quad
e^{\mu\nu}_{\times}=\cos 2\varphi_0 \hat{e}^{\mu\nu}_{\times}-\sin 2\varphi_0\hat{e}^{\mu\nu}_{+},\nb\\
e^{\mu\nu}_{l,2}&=& \sin\varphi_0\hat e^{\mu\nu}_{l,3}+\cos\varphi_0\hat e^{\mu \nu}_{l,2},\quad
e^{\mu\nu}_{l,3}=\cos\varphi_0\hat e^{\mu\nu}_{l,3}-\sin\varphi_0\hat e^{\mu \nu}_{l,2},\quad
e^{\mu\nu}_{o}= \hat{e}^{\mu\nu}_{o}.
\eqn
In \cite{Wang91}, the angle $\varphi_0$ defined in Eq.(\ref{varphi0}) was referred to as the polarization angle of the  GW with respect to the
($e_2, e_3$)-frame.  Note that the definition of the polarization angle is gauge invariant as shown explicitly in Appendix \ref{appda}.
 In addition, to the first-order of $h_\sigma$, this frame is  of parallel transport along the null geodesics defined by $l^{\mu}$ (See Appendix \ref{appdb1} for details),
\bq
l^{\nu}\nabla_{\nu}e^{\mu}_2 \simeq 0, \quad l^{\nu}\nabla_{\nu}e^{\mu}_3  \simeq 0.
\eq

Similarly, if we consider the null geodesic congruence defined by $n^{\mu}$,   we will find \cite{Wang91},
\bqn
\lb{geodesic2}
\frac{\mathrm{d}^2\zeta^{\mu}}{\mathrm{d}\lambda^2} &=& - R^{\mu}_{\nu\lambda\beta} n^{\nu}\eta^{\lambda}n^{\beta}\nb\\
&=&\Big\{\Big[(\Psi_2+\bar{\Psi}_2)-2\Phi_{11}+2\Lambda\Big]{(e^{\mu}_0+e^{\nu}_1)(e^{\nu}_0+e^{\mu}_1)}\nb\\
&&~{-}\Phi_{22} e^{\mu\nu}_{o}  +{\mbox{Re}} \Psi_4\; e^{\mu\nu}_{+}   {-}  {\mbox{Im}} \Psi_4\; e^{\mu\nu}_{\times}\nb\\
&&~{+} \sqrt{2}\Big[{\mbox{Re}}\left(\Phi_{12}+ \Psi_3\right) e^{\mu\nu}_{n,2}   +{\mbox{Im}}  \left(\Phi_{12}  {-} \Psi_3\right)e^{\mu\nu}_{n,3}\Big]\Big\} \eta_{\nu}, \nb\\
\eqn
where
\bqn
e^{\mu\nu}_{n,2} &\equiv& e^{\mu}_2 n^{\nu} + e^{\nu}_2 n^{\mu},\quad
e^{\mu\nu}_{n,3} \equiv e^{\mu}_{3}n^{\nu} + e^{\nu}_{3}n^{\mu},
\eqn
with $\zeta^{\mu}n_{\mu} = 0$. If we make rotation of the kind given by Eq.(\ref{rotation1}) but now with an angle $\varphi_4$,
\bq
\lb{varphi4}
\tan 2\varphi_4 \equiv  {+} \frac{{\mbox{Im}}\left(\Psi_4\right)}{{\mbox{Re}}\left(\Psi_4\right)},
\eq	
we shall obtain an expression similar to Eq.(\ref{varphi0}), and in particular we have
\bq
{\mbox{Re}} \Psi_4\; e^{\mu\nu}_{+}   {-}  {\mbox{Im}} \Psi_4\; e^{\mu\nu}_{\times} =  \left(\Psi_4\bar{\Psi}_4\right)^{1/2} \hat e^{\mu\nu}_{+}.
\eq
 Thus, $\varphi_4$ defines the polarization angle of the $\Psi_4$  GW,
moving in the opposite direction of $\Psi_0$. It can be shown that the ($e_2, e_3$)-frame is also  of parallel transport along the null
geodesics defined by $n^{\mu}$, so a such  defined polarization angle has an invariant interpretation.
The definition of the rotation angle in the FLRW background is consistent with the one in the Minkowski background \cite{Wang91},
 and it is independent of the nature of the geodesic deviations, null or timelike, as shown in Appendix \ref{appdb}.

Note that when $h_{\times} = 0$, we have ${\cal{P}}_{\times}(v) = 0$, and
\bq
{\mbox{Im}}\left(\Psi_0\right) = 0 = {\mbox{Im}}\left(\Psi_4\right), (h_{\times} = 0),
\eq
that is,  the polarization angles of $\Psi_0$ and $\Psi_4$ become zero, and all along the $e_2$-direction.

In addition, in the Minkowski background, we have ${\cal{H}} = 0$,  $\calhp = 0$ and $a = 1$, Then from Eqs.(\ref{RePsi0}) and (\ref{ImPsi0}),   we find that
\bqn
{\mbox{Re}}\left(\Psi_0\right) &=& {-}\frac{\calppp}{ \calr}\epsilon +O(\epsilon^2),\nb\\
{\mbox{Im}}\left(\Psi_0\right) &=& {+}\frac{\calptpp}{ \calr}\epsilon + O(\epsilon^2).
\eqn
Hence the polarization angle is given by
\bqn
\varphi_0^{(M_4)} &=& \frac{1}{2}\tan^{-1}\left(\frac{\calptpp}{\calppp}\right).
\eqn
Thus, along the path of the GW wave, we have
\bq
\frac{\partial\left(\varphi_{0}^{(M_4)}(v)\right)}{\partial u} = 0, \label{min}
\eq
 that is, the polarization of the wave does not change along its path in the Minkowski background, as it is usually expected.

 \subsection{Rotation of Polarization Angles in Curved Universe}

However,  the polarization angles $\varphi_{0}$ and $\varphi_{4}$ will change with time along their wave paths once the background is curved, as can be seen from
Eqs.(\ref{varphi0}) and (\ref{varphi4}). To study the rotations in more details, let us first introduce the ``scale-invariant" quantities via the relations \cite{WThesis,Szekeres72,Griffiths76a},
\bqn
\lb{eq2.31}
\Psi_0^{(0)} &\equiv& B^{-2}\Psi_0,
\quad \Phi_{02}^{(0)} \equiv (AB)^{-1}\Phi_{02},\nb\\
\Phi_{00}^{(0)} &\equiv& B^{-2}\Phi_{00}, \quad \Phi_{11}^{(0)} \equiv (AB)^{-1}\Phi_{11},
\eqn
we find that
\bqn
\lb{eq2.32a}
{\left[\text{Re}\left(\Psi_0^{(0)}\right)\right]}_{,u} &=& \frac{1}{2}\Bigg\{U_{,u} \left[\text{Re}\left(\Psi_0^{(0)}\right)\right]
 - U_{,v} \text{Re}\left(\Phi_{02}^{(0)}\right)
- 2 {\left[\text{Re}\left(\Phi_{02}^{(0)}\right)\right]}_{,v} \nb\\
&& ~~~- \epsilon\left(2h_{+, v} \Phi_{11}^{(0)} + h_{+, u} \Phi_{00}^{(0)}\right) \Bigg\}
 + {\cal{O}}\left(\epsilon^2\right),\\
\lb{eq2.32b}
{\left[\text{Im}\left(\Psi_0^{(0)}\right)\right]}_{,u} &=& \frac{1}{2}\Bigg\{U_{,u} \left[\text{Im}\left(\Psi_0^{(0)}\right)\right]
 - U_{,v} \text{Im}\left(\Phi_{02}^{(0)}\right)
  - 2 {\left[\text{Im}\left(\Phi_{02}^{(0)}\right)\right]}_{,v} \nb\\
 && ~~~ +  \epsilon\left(2h_{\times, v} \Phi_{11}^{(0)} + h_{\times, u} \Phi_{00}^{(0)}\right)\Bigg\}
 + {\cal{O}}\left(\epsilon^2\right),
\eqn
where $\Psi_0$ is given by Eq.(\ref{A1}), and
\bqn
\lb{eq2.32c}
\text{Re}\left(\Phi_{02}^{(0)}\right) &=&  \frac{1}{2}\Big(h_{+,\eta\eta} - h_{+,zz} + 2{\cal{H}}h_{+,\eta}\Big)\epsilon+ {\cal{O}}\left(\epsilon^2\right),\nb\\
\text{Im}\left(\Phi_{02}^{(0)}\right) &=& - \frac{1}{2}\Big(h_{\times,\eta\eta} - h_{\times,zz} + 2{\cal{H}}h_{\times,\eta}\Big)\epsilon + {\cal{O}}\left(\epsilon^2\right),\nb\\
\Phi_{11}^{(0)} &=&   \frac{1}{2}\left(2{\cal{H}}^2  - \frac{a''}{a}\right)  + {\cal{O}}\left(\epsilon\right),\nb\\
\Phi_{00}^{(0)} &=&  2{\cal{H}}^2 - \frac{a''}{a}  + {\cal{O}}\left(\epsilon^2\right),
\eqn
as can be seen from   Eqs.(\ref{A2}) in Appendix \ref{appda}. From these expressions we find that
\bqn
\lb{eq2.33}
{\left[\text{Re}\left(\Psi_0^{(0)}\right)\right]}_{,u} &\propto&   f\left(h_{+, a}, h_{+, ab}\right)\epsilon + {\cal{O}}\left(\epsilon^2\right), \nb\\
{\left[\text{Im}\left(\Psi_0^{(0)}\right)\right]}_{,u} &\propto&  g\left(h_{\times, a}, h_{\times, ab}\right)\epsilon + {\cal{O}}\left(\epsilon^2\right),
\eqn
which implies that the ``$\times$" modes cannot be created from the ``$+$" modes, or vice versa, where $a, b = u, v$ and $f, g$ are certain functions whose explicit forms are not important.
To be more specific, let us assume that $h_{+} \not= 0, h_{\times} = 0$ at a given moment, say, $t =t_0$. Then, since the function  $g$ does not depend on $h_{+}$, it vanishes
at $t = t_0$, too. As a result, the imaginary part of $\Psi_0^{(0)}$ shall remain constant along the wave path, as one can see from Eq.(\ref{eq2.33}). On the other hand, if the
function $g$ depends on $h_{+}$,  in general it will not vanish  at  $t_0$ even when $h_{\times} = 0$,
as now $h_{+} \not= 0$. So,  the imaginary part of $\Psi_0^{(0)}$  now can no longer remain constant along the  wave path, and the polarization angle defined above will change, too.
As shown explicitly in Appendix B, this is possible only when the second-order effects are taken into account, and to the linear order of $\epsilon$ such effects do not exist.
For more general cases, see Eqs.(4.10) and (4.11) given in \cite{Wang91}.

However, due to the presence of the matter field of the FLRW background,
represented by   $\Phi_{00}^{(0)}$ and $\Phi_{11}^{(0)}$,  as well as its linear perturbations, represented by $\Phi_{02}^{(0)}$, both of the ``$\times$" and ``$+$" modes can get  amplified or diluted,
depending on their amplitudes and signs, so in principle
the effects  to   the ``$\times$" modes are different from those to the ``$+$" modes. As a result, the polarization angle $\varphi_{0}$ defined above will change along the wave path.
The above analysis was for the GW represented by $\Psi_0$.

A similar analysis can be carried out for the $\Psi_4$ wave, too, and the same  conclusion will be obtained, that is, the ``$\times$" modes cannot be created from the ``$+$" modes, or vice versa, but
the polarization angle of the  $\Psi_4$ wave can still get changed, due to the different effects of the matter presented in our universe to the two different kinds of polarization modes.

Therefore,  to the linear order of $\epsilon$, there is no transfer between the two different kinds of modes.
However,   due to the fact that the effects of  the curved background on each of the two independent modes depend on their waveforms, the polarization angle can be still changed along the wave path.
These can be seen more clearly from the following studies of the polarization angles in various epochs of the universe.

Before proceeding to the next section, it should be emphasized that the conclusion that there is no transfer between the two different modes holds only in the linear level (of $\epsilon$), as shown explicitly in \cite{Wang91,WThesis} that the nonlinear  interaction of the GW wave with the background will  lead to the phenomena of the Faraday rotation.

\section{Propagation of Gravitational Waves in Different Epochs of the Universe}\label{3}
\renewcommand{\theequation}{3.\arabic{equation}} \setcounter{equation}{0}

To study the propagation and polarization of the GWs further, let us turn to consider different epochs  in which the universe is dominated by different components of matter fields.
To show explicitly the dependence of the polarization on the background,   in this section we first assume that there is only one epoch throughout the whole cosmic
history: radiation-dominated, matter-dominated or de sitter epoch, respectively. Then, we shall combine these epochs together within the framework of the $\Lambda$CDM model.  It should be noticed that the expression of $\varphi_0$ in Eq.(\ref{varphi0}) (or $\varphi_4$ in Eq.(\ref{varphi4}) ) holds for any GW sources. However, in order to have more Intuitive understanding, let us consider a binary system as a concrete example.

\subsection{Radiation-Dominated Epoch}\label{3.1}

The scale factor in general  is given by a power law  $a =({\eta}/{\eta_0})^p$  in terms of the conformal time. In particular,
in the radiation-dominated epoch we have $p=1$. Then, the tail vanishes ${\cal{T}} = 0$ \cite{Chu15}, and from Eqs.(\ref{RePsi0}) and (\ref{ImPsi0}) together with $z \gg x, y$, we find that
\begin{eqnarray}
{\mbox{Re}}\left(\Psi_0\right) &=& {-} \frac{1}{4A^2a^4 \calr}\Bigg(\calppp -\frac{\sqrt{2}}{\eta}\calpp + \frac1{\eta^2}\calp\Bigg)\epsilon +{\cal{O}}\left(\epsilon^2\right), \nb \\
{\mbox{Im}}\left(\Psi_0\right) &=&\frac{1}{4A^2a^4 \calr}\Bigg(\calptpp -\frac{\sqrt{2}}{\eta}\calptp  + \frac1{\eta^2}\calpt \Bigg)\epsilon +{\cal{O}}\left(\epsilon^2\right).
\end{eqnarray}

Substituting the above equations into Eq.(\ref{varphi0}) and then performing  the series expansion in terms of $\eta$, we find that
\begin{eqnarray}
\lb{phi}
\varphi_0 (u, v) &=&\frac1{2}\arctan\left(\frac{\calptpp}{\calppp}\right)+ \frac{ g(v)}{\sqrt{2}\; \eta}, \nb\\
\varphi_{0}(u, v)_{,u} &=& -\frac{g(v)}{2\eta^2},
\end{eqnarray}
where
\bqn
\lb{gv}
g(v)&\equiv& \frac{\calpp \calptpp-\calptp \calppp}{\calppp^2+\calptpp^2}
\simeq \frac{1+z_s}{2\sqrt{2}\pi f_{gw}[\hat{\tau}(v)]}.
\eqn
In the second step of Eq.(\ref{gv}), we have substituted the result of ${\cal{P}}_{\sigma}$ for the binary system, which can be found explicitly in Appendix \ref{appdc}. And it is obvious that $g(v)$ only depends on GWs frequency $f_{gw}[\hat{\tau}(v)]$ at the emission time.

\begin{figure}[htb]
\centering
\includegraphics[scale=0.5]{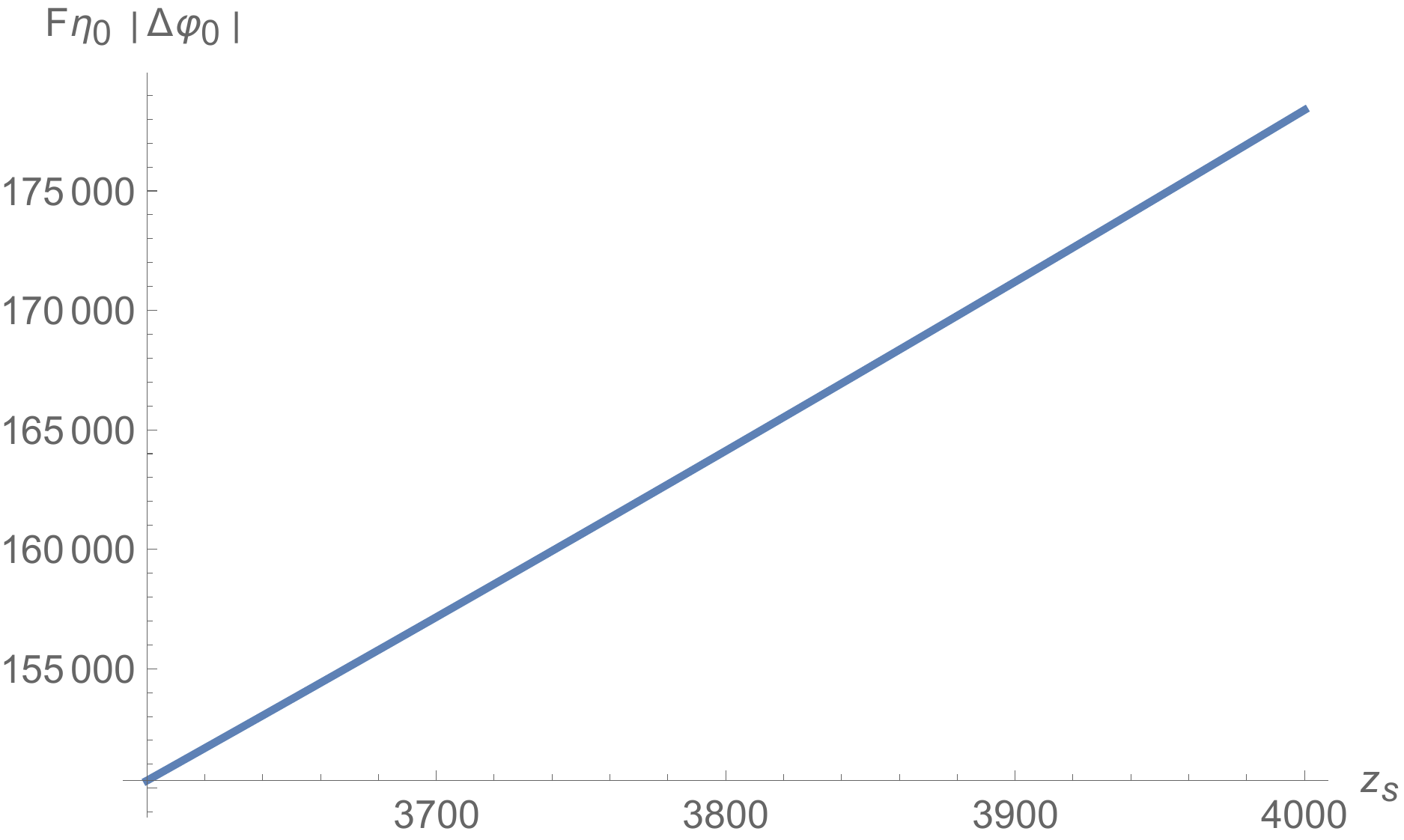}
\caption{The polarization angle $\left| \Delta \varphi_0 \right|$ is a function of the redshift
 $z_s$ in the radiation-dominated epoch, and we set $3600\leq z_s\leq 4000$. } \label{fig:GWrotation1}
\end{figure}

Suppose that the radiation-dominated epoch is through the whole cosmic history, so that $\eta_0$ is the age of the universe, then the change of the polarization
 angle from a moment $\eta_e$  to the Earth along the wave path ($v = $ Constant)  can be obtained from Eq.(\ref{phi}), which is given by \footnote {This can be also
 obtained by the following considerations:  Let us consider two observers, and without loss of the generality, we assume that they are static and located on the ($\eta, z$)-plane,
one is referred to as $O_e$ and locates at $z = z_e$,  and the other is referred to as $O_o$  and locates at the origin $z = z_0 = 0$. Assume that at the moment $\eta = \eta_e$, a GW passes $O_e$
with its polarization angle $\left. \varphi_0 (u, v)\right|_{(\eta_e, z_e)}$. At the moment $\eta_0$, it passes by $O_o$ with  its polarization angle $\left. \varphi_0 (u, v)\right|_{(\eta_0, z_0)}$. Then,
 from Eq.(\ref{gv}) and Fig. \ref{fig:Ptz}, it can be seen that the difference between these two polarization angle is given by Eq.(\ref{Dphi}), too.},
\begin{eqnarray}
\lb{Dphi}
\mid \Delta \varphi_0 \mid
 &=&\left|\frac1{\sqrt{2}}\Big(\frac1{\eta_0}-\frac1{\eta_e}\Big)\; g(v_e)\right|\nb\\
&=&\left|\frac1{\sqrt{2}\eta_0}\Big(1-\frac1{a(\eta_e)}\Big)\; g(v_e)\right|,
\end{eqnarray}
where  $v_e$ is the corresponding retarded time, which is a constant along the GW path,  and $a(\eta_0)=1$, $a(\eta_e)=1/(1+z_s)$. Then, we find that
\begin{eqnarray}
 \left| \Delta \varphi_0 \right| &=&\left|-\frac1{\sqrt{2}}\frac{z_s}{\eta_0}\; g(v_e)\right|
\simeq\frac1{4}\cdot\frac{z_s}{\eta_0}\frac{(1+z_s)}{\pi f_{gw}(\hat{\tau})}.
\end{eqnarray}
Thus, in the   source frame, we have
\bq
\left| \Delta \varphi_0 \right|  \simeq \frac1{4}\frac{z_s}{\eta_0 }\frac{(1+z_s)}{\pi f^{(s)}_{gw}(\hat{\tau}_s)},
\eq
with
\bq
f^{(s)}_{gw}(\hat{\tau}_s)=\frac1{\pi}\Big(\frac{5}{256}\frac{1+z_s}{\hat{\tau}_s}\Big)^{3/8}\left(G_NM_c\right)^{-5/8},
\eq
where $\tau_s$ denotes the time to the coalescence measured by the source's clock, and $f^{(s)}_{gw}{(\tau_s)}$ is the frequency in the source frame.
However, it is more convenient to express the polarization angle in the observer's frame, which is given by
\bqn
\lb{rotation r}
\left| \Delta \varphi_0 \right|  &\simeq&  \frac1{4}\frac{z_s}{\eta_0 }\frac{(1+z_s)}{\pi f^{(obs)}_{gw}(\hat{\tau}_{obs})(1+z_s)},\nb\\
 &\simeq&  \frac1{4}\frac{z_s}{\eta_0}\frac{(1+z_s)^{5/8}}{ F(\tau_{obs})},
\eqn
with
\begin{eqnarray}
f^{(obs)}_{gw}(\hat{\tau}_{obs})&=&\frac1{1+z_s}f^{(s)}_{gw}(\hat{\tau}_s), \quad \hat{\tau}_{obs}=(1+z_s)\hat{\tau}_s,\nb\\
F(\tau_{obs})&\equiv&\Big(\frac{5}{256}\frac{1}{\tau_{obs}}\Big)^{3/8}\left(G_NM_c\right)^{-5/8},
\end{eqnarray}
where $f^{(obs)}_{gw}(\tau_{obs})$ and $\tau_{obs}$ are the frequency and time to the coalescence in the observer's frame.
In Fig. \ref{fig:GWrotation1}, we plot $\left|\Delta \varphi_0 \right| $ as a function of $z_s$.

\subsection{Matter-Dominated Epoch}\label{3.2}

In the matter-dominated epoch, we have  $a =({\eta}/{\eta_0})^2$ and,
\begin{eqnarray}
D_{ij}^{(tail)} &=& \frac{4G_N}{\eta_0^2a^{3/2}(\eta)}\hat{D}_{ij}^{(tail)}(v),\nb\\
\hat{D}_{ij}^{(tail)}(v) &=&      \int_{0}^{v-0^{+}}{\mathrm{d}\eta'a^{5/2}}(\eta')
 {\int_{\mathbb{R}^3}{\mathrm{d}^3\vec{x}'   \Pi_{ij}\left(\eta', \vec{x}'\right)}}.\nb\\
\end{eqnarray}
Then, from Eq.(\ref{2.2}) we find that
\begin{eqnarray}
{\cal{T}}_\sigma(u, v) = \frac{{\cal{C}}_\sigma(v)}{a^{1/2}(\eta)},
\end{eqnarray}
where ${\cal{C}}_\sigma(v)$ is related to the source as $ {\cal{C}}_\sigma(v)\simeq (G_N/\eta_0^2) \hat{D}_{ij}^{(tail)}(v)$. Hence, substituting $n = \frac{3}{2}$ and in the limit $x,y << z$ we obtain
\begin{eqnarray}
{\mbox{Re}}\left(\Psi_0\right) &=& -\frac{1}{4A^2a^4}\Bigg[\frac{1}{\calr}\Bigg(\calppp - \frac{2\sqrt{2}}{\eta}\calpp+\frac{3}{\eta^2}\calp \Bigg)\nb\\
&&~~+\frac{1}{a^{2}}\Bigg({\cal{C}}_{+}(v)_{,vv}\frac{4\sqrt{2}}{\eta}{\cal{C}}_{+}(v)_{,v}+\frac{19}{2\eta^2}{\cal{C}}_{+}(v)\Bigg)\Bigg]\epsilon  + {\cal{O}}\left(\epsilon^2\right), \nb\\
{\mbox{Im}}\left(\Psi_0\right) &=&\frac{1}{4A^2a^4}\Bigg[\frac{1}{\calr}\Bigg(\calptpp - \frac{2\sqrt{2}}{\eta}\calptp+\frac{3}{\eta^2}\calpt \Bigg)\nb\\
&&~~+\frac{1}{a^{2}}\Bigg({\cal{C}}_{\times}(v)_{,vv}\frac{4\sqrt{2}}{\eta}{\cal{C}}_{\times}(v)_{,v}+\frac{19}{2\eta^2}{\cal{C}}_{\times}(v)\Bigg)\Bigg]\epsilon+{\cal{O}}\left(\epsilon^2\right).
\end{eqnarray}

It is obvious that ${\cal{C}}(v)$ is highly suppressed by ${P_{\sigma}(v)}/{\calr}$  while taking $\eta_0$ as the age of the universe, we find
\bq
\lb{CP}
\left| \frac{{\cal{C}}_\sigma(v_0)}{P_{\sigma}(v_0)/\calr}\right| \simeq \frac{\Delta t}{\eta_0} \frac{\left| \vec{x} \right|}{\eta_0}a^{-3/2}(\eta_{\ast}),
\eq
where $a(\eta_0)=1$, $\Delta t \simeq \int_{peak width}{d\eta a(\eta)}$, and $\eta_{\ast}$ is the peak time of the source's strength. Then,  we find
\begin{eqnarray}
\varphi_0 (u, v)&=& \frac1{2}\arctan\left(\frac{\calptpp}{\calppp}\right)+\frac{\sqrt{2}}{\eta}\cdot g(v)+{\cal{O}}\left(\frac1{\eta^2}\right),  \nb \\
\varphi_{0}(u, v)_{,u}&=& -\frac{1}{\eta^2}\cdot g(v)+{\cal{O}}\left(\frac1{\eta^3}\right).
\end{eqnarray}

\begin{figure}[htb]
\centering
\includegraphics[scale=0.5]{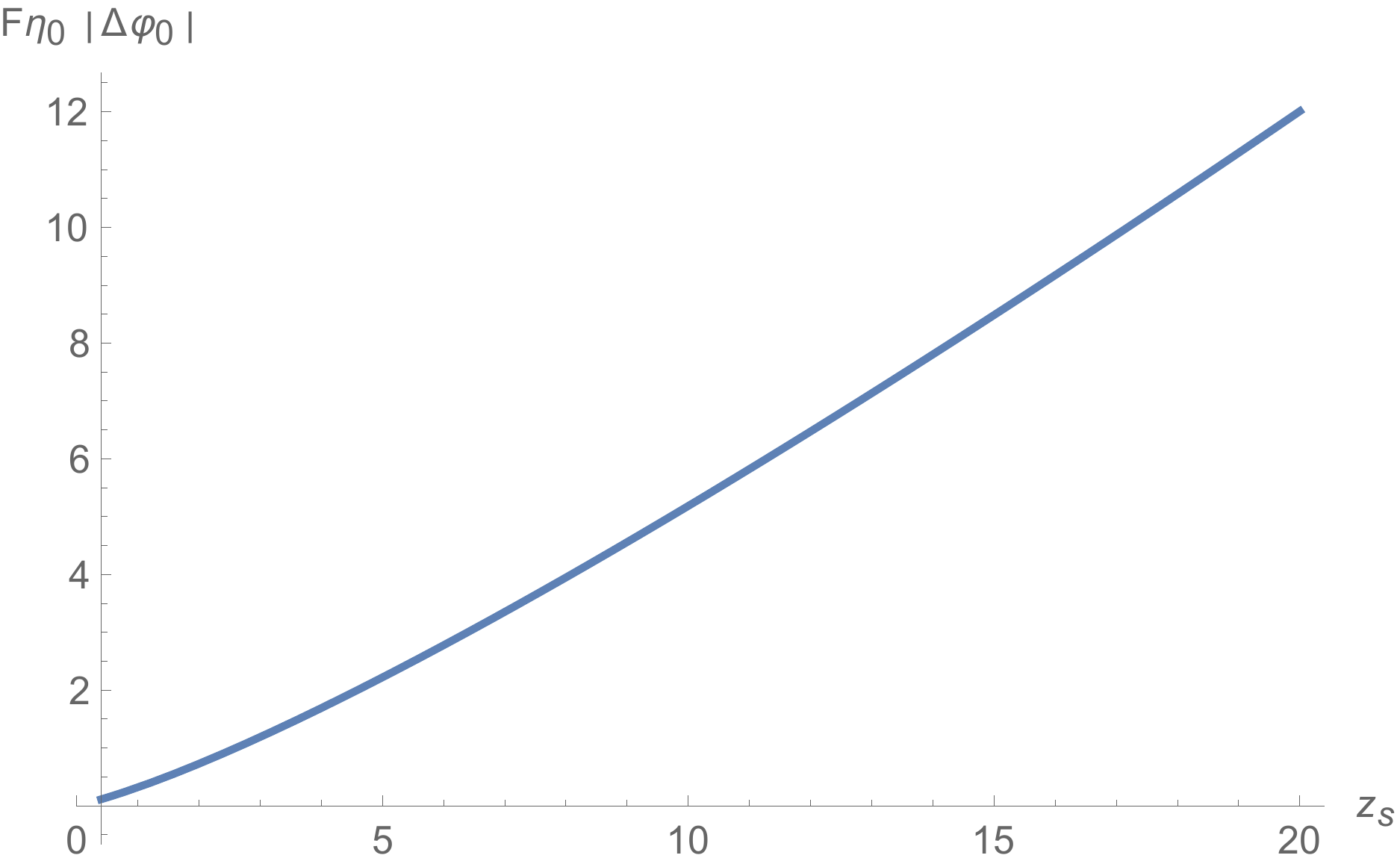}
\caption{The polarization rotation $\left| \Delta \varphi_0 \right|$ is a function of the redshift
 $z_s$ in the matter-dominated epoch, and we set $0.4\leq z_s\leq 20$ . } \label{fig:GWrotation2}
\end{figure}

  Assume that the matter-dominated epoch is through the whole cosmic history, then the accumulation of the polarization angle from the source to the earth is given by,
\begin{eqnarray}
\left| \Delta \varphi_0 \right| &\simeq&\left|\sqrt{2} \left(\frac1{\eta_0}-\frac1{\eta_e}\right)\cdot g(v_e)\right|\nb\\
&\simeq&\left|\frac{\sqrt{2}} {\eta_0}\left(1-\frac1{\sqrt{a(\eta_e)}}\right)\cdot g(v_e)\right|\nb\\
&\simeq&\left|\frac{\sqrt{2}}  {\eta_0}\left(1-\sqrt{1+z_s}\right )\cdot g(v_e)\right|.\nb\\
\end{eqnarray}
Thus, the  rotation angle in observer's frame is
\begin{eqnarray}
\left|\Delta \varphi_0\right| &\simeq&\left|\frac1{2\eta_0}\left(1-\sqrt{1+z_s}\right)\cdot\frac{(1+z_s)^{5/8}}{ F(\tau_{obs})}\right|.\nb\\
\end{eqnarray}
 We can also plot $\left|\Delta \varphi_0 \right| $ as a function of $z_s$, which is given in Fig. \ref{fig:GWrotation2}.

\subsection{de Sitter Background}\label{3.3}

In this case, the tail $D_{ij}^{(tail)}(\eta, \vec{x})$ takes the form \cite{Chu15},
\bqn
\lb{eq3.26}
D_{ij}^{(tail)}(\eta, \vec{x}) &=& \frac{4G_N}{\eta_0^2} \hat{D}_{ij}^{(tail)}(v),\nb\\
\hat{D}_{ij}^{(tail)}(v) &\equiv& \int_{-\infty}^{v -0^{+}}{\mathrm{d}\eta' a^{4}(\eta')}\int_{\mathbb{R}^3}{\mathrm{d}^3\vec{x}'   \Pi_{ij}\left(\eta', \vec{x}'\right)}.\nb\\
\eqn
In the present case, we have
\bqn
\lb{eq3.27}
{\cal{T}}_\sigma = a(\eta) {\cal{D}}_\sigma(v),
\eqn
where ${\cal{D}}_\sigma(v) \simeq  {(G_N/\eta_0^2)} \hat{D}_{ij}^{(tail)}(v)$  is  related to the integration of the source, as that given by Eq.(\ref{eq3.26}). With $n=0$ and $a(\eta)= ({\eta_0}/{\eta})$,  from Eqs.(\ref{RePsi0}) and (\ref{ImPsi0}),
we find that
\bqn
\lb{eq3.28}
{\mbox{Re}}\left(\Psi_0\right) &=& -\frac1{4A^2a^4}\Bigg[\frac{1}{\calr}\Bigg(\calppp +\frac{\sqrt{2}}{\eta}\calpp\Bigg)\nb\\
&&~~+a\Bigg({\cal{D}}_{+}(v)_{,vv}-\frac{\sqrt{2}}{\eta}{\cal{D}}_{+}(v)_{,v}+\frac{1}{2\eta^2}{\cal{D}}_{+}(v)\Bigg)\Bigg]\epsilon  + {\cal{O}}\left(\epsilon^2\right), \nb\\
{\mbox{Im}}\left(\Psi_0\right) &=&  \frac1{4A^2a^4}\Bigg[\frac{1}{\calr}\Bigg(\calptpp +\frac{\sqrt{2}}{\eta}\calptp\Bigg)\nb\\
&&~~+a \Bigg({\cal{D}}_{\times}(v)_{,vv}-\frac{\sqrt{2}}{\eta}{\cal{D}}_{\times}(v)_{,v}+\frac{1}{2\eta^2}{\cal{D}}_{\times}(v)\Bigg)\Bigg]\epsilon + {\cal{O}}\left(\epsilon^2\right).
\eqn

Similar to   the matter-dominated case, now ${\cal{D}}_\sigma(v)$ is  bounded by
\begin{eqnarray}
\left| {\cal{D}}_\sigma(v)\right|  \lesssim  {(G_N/\eta_0^2)} \Delta t\left| \int_{\mathbb{R}^3}{\mathrm{d}^3\vec{x}'  a^{3}(\eta_{\ast})\Pi_{ij}\left(\eta_{\ast}, \vec{x}'\right)}\right|,\nb\\
\end{eqnarray}
where $\Delta t \simeq \int_{peak width}{d\eta a(\eta)}$, and $\eta_{\ast}$ is the peak time of the source's strength. Then, we find that,
\bq
\lb{DP}
\left| \frac{{\cal{D}}_\sigma(v)}{P_{\sigma}(v)/\calr}\right|  \sim  {(H\cdot\Delta t)\cdot( H\cdot a(\eta)\left| \vec{x}\right|)},\\
\eq
where $\eta_{0}=-\frac1{H}$. Substituting Eq.(\ref{eq3.28}) into Eq.(\ref{varphi0}) and then carrying out the series expansion, we finally find
\begin{eqnarray}
\varphi_{0}(u, v) &=&\frac1{2}\arctan\left(\frac{\calptpp}{\calppp}\right)- \frac{1}{\sqrt{2}\eta}\cdot g(v)+O(\frac1{\eta^2})\nb \\
\varphi_{0}(u, v)_{,u}&=&\frac{1}{2\eta^2}\cdot g(v) +{\cal{O}}\left(\frac1{\eta^3}\right).
\end{eqnarray}

\begin{figure}[htb]
\centering
\includegraphics[scale=0.5]{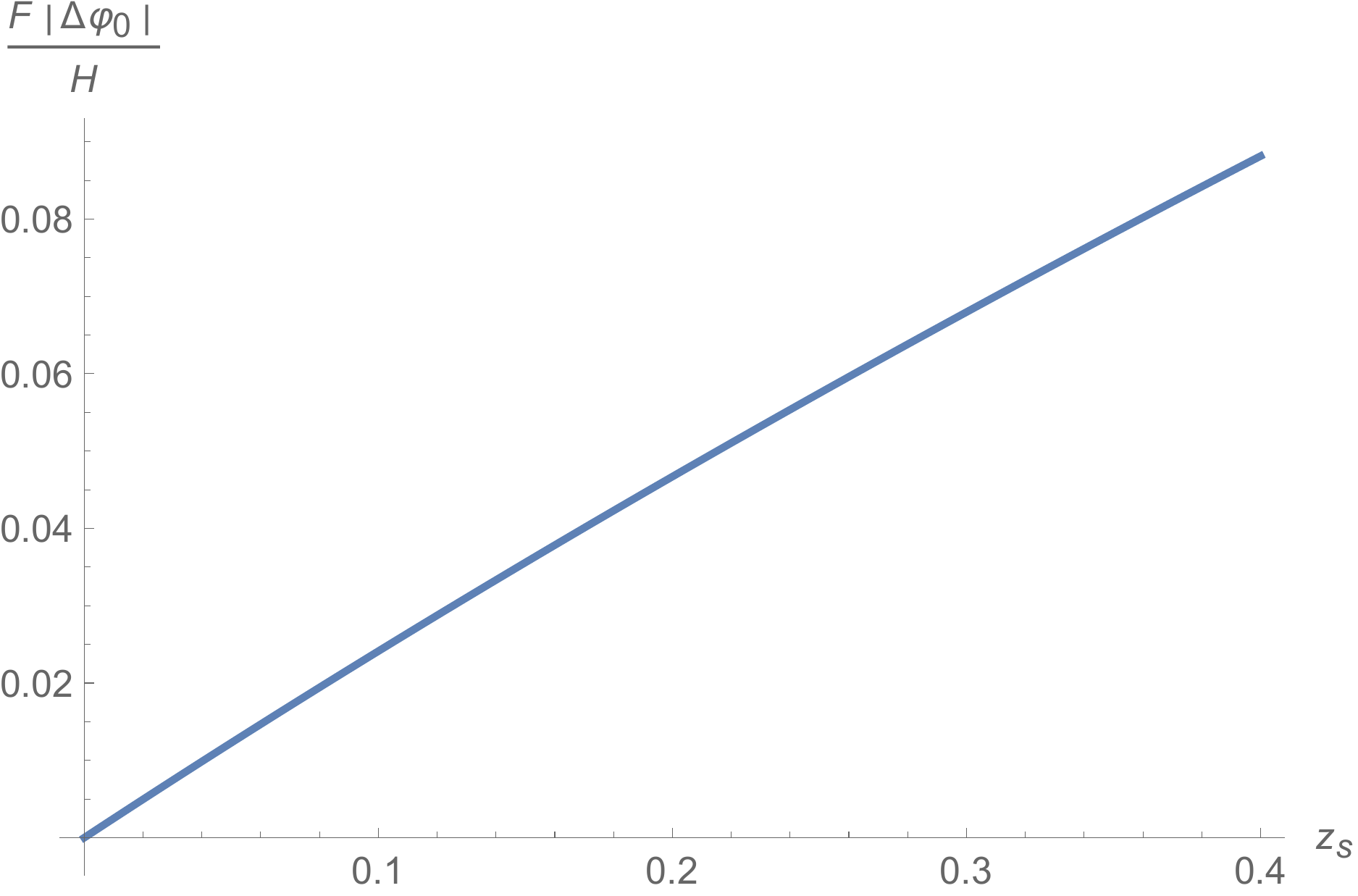}
\caption{The polarization rotation angle $\left| \Delta \varphi_0 \right|$ is a function of the redshift
 $z_s$ in de sitter epoch, and we set $0\leq z_s\leq 0.4$.} \label{fig:GWrotation3}
\end{figure}

Similarly, supposing that the de sitter epoch is through the whole cosmic history, we have
\begin{eqnarray}
\left| \Delta \varphi_0 \right|  &\simeq&\left| \frac1{\sqrt{2}}\Big(\frac1{\eta_0}-\frac1{\eta_e}\Big)\cdot g(v_e)\right| \nb\\
&\simeq&\left|\frac1{\sqrt{2}\eta_0}\Big(1-a(\eta_e)\Big)\cdot g(v_e)\right| \nb\\
&\simeq&\left|\frac1{\sqrt{2}\eta_0}\Big(1-\frac1{1+z_s}\Big)\cdot g(v_e)\right|.\nb\\
\end{eqnarray}
Thus,  the polarization angle in observer's frame can be expressed as,
\begin{eqnarray}
\left| \Delta \varphi_0 \right| &\simeq&\frac{H}{4}\left(\frac{z_s}{1+z_s}\right)\cdot\frac{(1+z_s)^{5/8}}{F(\tau_{obs})}.
\end{eqnarray}
To see this effect more explicitly, we  plot $\left| \Delta \varphi_0 \right| $ as a function of $z_s$  in Fig. \ref{fig:GWrotation3}.

 \subsection{More realistic considerations}

\begin{figure}[htb]
\centering
\includegraphics[scale=0.45]{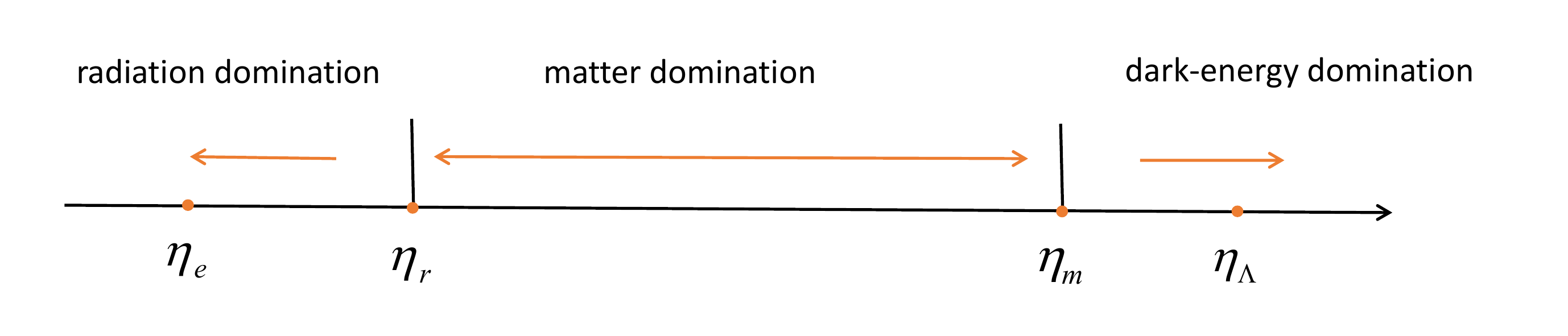}
\caption{Sketch of propagations of GWs in the evoluated universe. $\eta_e$ is the emission time of the GWs, $\eta_r$ denotes the time when radiation-domination is ended, $\eta_m$ denotes the time when matter-domination epoch is ended, and $\eta_{\Lambda}$ is the receiving time of the detector.} \label{fig:P1}
\end{figure}

In the last subsections, we have assumed that there is only one epoch throughout the whole cosmic history. The real universe is obviously not the case. In this subsection, we will consider a more realistic case by combining all these epochs
together within the framework of the $\Lambda$CDM model.  We consider a binary system emitted GW signals at some time $\eta_e$. Then, we should integrate  the polarization rotations from $\eta_e$ to $\eta_{\Lambda}$, the receiving time of detectors. The whole picture is sketched in Fig. \ref{fig:P1}.

Obviously,  we have
\begin{eqnarray}
\eta_e &=&\int^{z_s}_{\infty} \frac{-\mathrm{d}z}{H_0\sqrt{\Omega_{\Lambda}+\Omega_M(1+z)^3+\Omega_R(1+z)^4}},\nb\\
\eta_r &=&\int^{z_r}_{\infty} \frac{-\mathrm{d}z}{H_0\sqrt{\Omega_{\Lambda}+\Omega_M(1+z)^3+\Omega_R(1+z)^4}},\nb\\
\eta_m &=&\int^{z_m}_{\infty} \frac{-\mathrm{d}z}{H_0\sqrt{\Omega_{\Lambda}+\Omega_M(1+z)^3+\Omega_R(1+z)^4}},\nb\\
\eta_{\Lambda} &=&\int^{0}_{\infty} \frac{-\mathrm{d}z}{H_0\sqrt{\Omega_{\Lambda}+\Omega_M(1+z)^3+\Omega_R(1+z)^4}}.\nb\\
\end{eqnarray}
The fucntion $g(v_e)$  in observer's frame is given by
\begin{eqnarray}
g(v_e)&\simeq&\frac1{2\sqrt{2}}\frac{(1+z_s)^{5/8}}{ F(\tau_{obs})}, \nb\\
F(\tau_{obs})&\equiv&\Big(\frac{5}{256}\frac{1}{\tau_{obs}}\Big)^{3/8}\left(G_NM_c\right)^{-5/8},\nb\\
\end{eqnarray}
where $z_s$ is the  redshift of the source at $\eta_e$, and $z_r\simeq 3600$ at $\eta_r$ \cite{Ryden:2003yy}, $z_m\simeq0.4$ at $\eta_m$  \cite{Ryden:2003yy}.  $\Omega_{\Lambda}$, $\Omega_M$ and $\Omega_R$ are the density parameters for dark energy, matter and radiation, respectively. $H_0$ is the Hubble constant.

Depending on the values of $z_s$, the total rotation angle of the whole propagation process can be divided into the following three cases.

\begin{itemize}

  \item $0<z_s<0.4$, that is,  $\eta_m<\eta_e<\eta_\Lambda$. The accumulated polarization angle from $\eta_e$ to $\eta_{\Lambda}$ is
  \begin{eqnarray}
\left|\Delta \varphi_0\right| &\equiv&\left|\Delta \varphi_0\right|_{\Lambda}.
\end{eqnarray}
with
\begin{eqnarray}
\left|\Delta \varphi_0\right|_{\Lambda} &\simeq&\left|\int^{\eta_{\Lambda}}_{\eta_e}\frac1{\sqrt{2}\eta^2}\mathrm{d}\eta \; g(v_e)\right|,
 \end{eqnarray}
where $\left|\Delta \varphi_0\right|_\Lambda$ is the polarization angle  from $\eta_e$ to $\eta_{\Lambda}$.

  \item  $0.4<z_s<3600$, that is,  $\eta_r<\eta_e<\eta_m$. The accumulated polarization angle from $\eta_e$ to $\eta_{\Lambda}$ is
  \begin{eqnarray}
\left|\Delta \varphi_0\right| &\equiv&\left|\Delta \varphi_0\right|_m+\left|\Delta \varphi_0\right|_{\Lambda}.
\end{eqnarray}
with
\begin{eqnarray}
\left|\Delta \varphi_0\right|_m &\simeq&\left|-\int^{\eta_m}_{\eta_e}\frac{\sqrt{2}}{\eta^2}\mathrm{d}\eta \; g(v_e)\right|,\nb\\
\left|\Delta \varphi_0\right|_{\Lambda} &\simeq&\left|\int^{\eta_{\Lambda}}_{\eta_m}\frac1{\sqrt{2}\eta^2}\mathrm{d}\eta \; g(v_e)\right|,
\end{eqnarray}
where  $\left|\Delta \varphi_0\right|_m$ is the polarization angle from  $\eta_e$ to $\eta_m$, and $\left|\Delta \varphi_0\right|_\Lambda$ is the rotation angle  from $\eta_m$ to $\eta_{\Lambda}$.

 \item $z_s>3600$, that is,  $\eta_e<\eta_r$. The accumulated polarization angle from $\eta_e$ to $\eta_{\Lambda}$ is
  \begin{eqnarray}
\left|\Delta \varphi_0\right| &\equiv&\left|\Delta \varphi_0\right|_r+\left|\Delta \varphi_0\right|_m+\left|\Delta \varphi_0\right|_{\Lambda}.\nb\\
\end{eqnarray}
with
\begin{eqnarray}
\left|\Delta \varphi_0\right|_r &\simeq&\left|-\frac1{\sqrt{2}}\int^{\eta_r}_{\eta_e}\frac1{\eta^2}\mathrm{d}\eta \; g(v_e)\right|,\nb\\
\left|\Delta \varphi_0\right|_m &\simeq&\left|-\int^{\eta_m}_{\eta_r}\frac{\sqrt{2}}{\eta^2}\mathrm{d}\eta \; g(v_e)\right|,\nb\\
\left|\Delta \varphi_0\right|_{\Lambda} &\simeq&\left|\int^{\eta_{\Lambda}}_{\eta_m}\frac1{\sqrt{2}\eta^2}\mathrm{d}\eta \; g(v_e)\right|,
\end{eqnarray}
where $\left|\Delta \varphi_0\right|_r$  is the polarization angle from  $\eta_e$ to $\eta_r$, $\left|\Delta \varphi_0\right|_m$ is the polarization angle from  $\eta_r$ to $\eta_m$, and $\left|\Delta \varphi_0\right|_\Lambda$ is the
polarization angle from $\eta_m$ to $\eta_{\Lambda}$.
\end{itemize}

\begin{figure}[htb]
\centering
\includegraphics[scale=0.5]{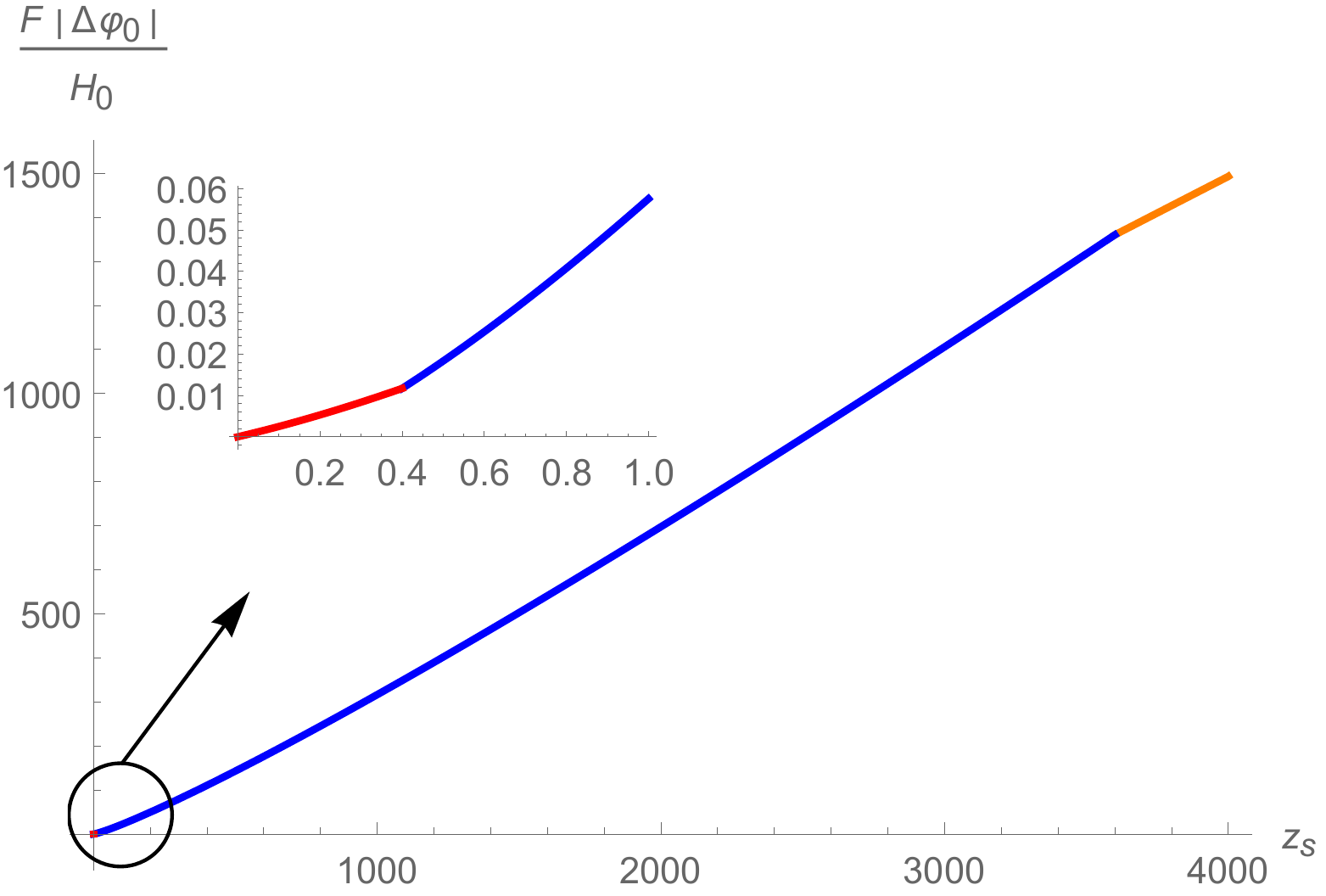}
\caption{The polarization rotation angle $\frac{F\left| \Delta \varphi_0 \right|}{H_0}$ as a function of the redshift
 $z_s$ in the whole propagation process. The red line is the first section when $0<z_s<0.4$, the blue line is the second section when $0.4<z_s<3600$, and the orange line is the third section when $3600<z_s<4000$. } \label{fig:GWrotation4}
\end{figure}

\begin{figure*}[htb]
\centering
\includegraphics[width=0.3\textwidth]{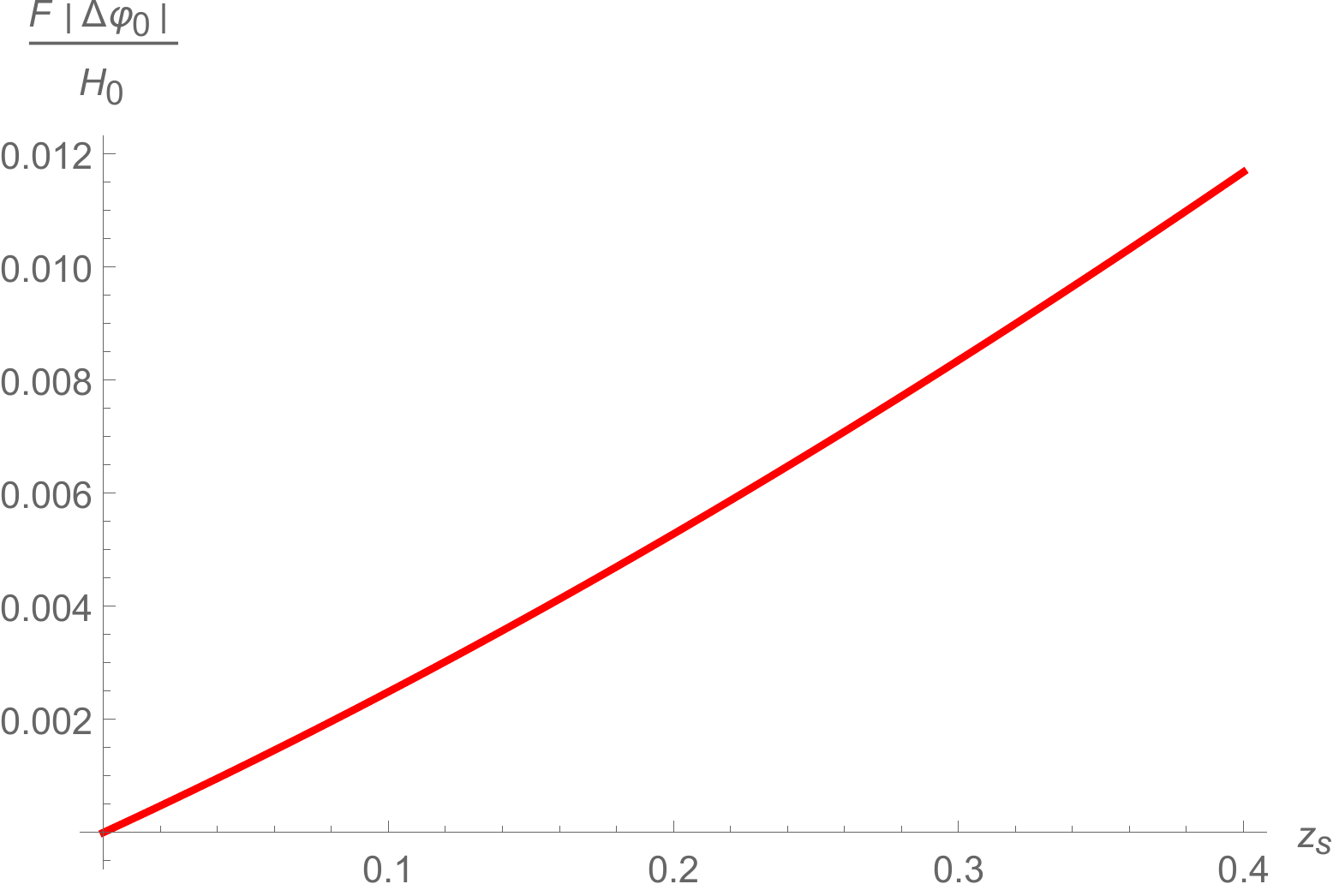}\hspace{0.7cm}
\includegraphics[width=0.3\textwidth]{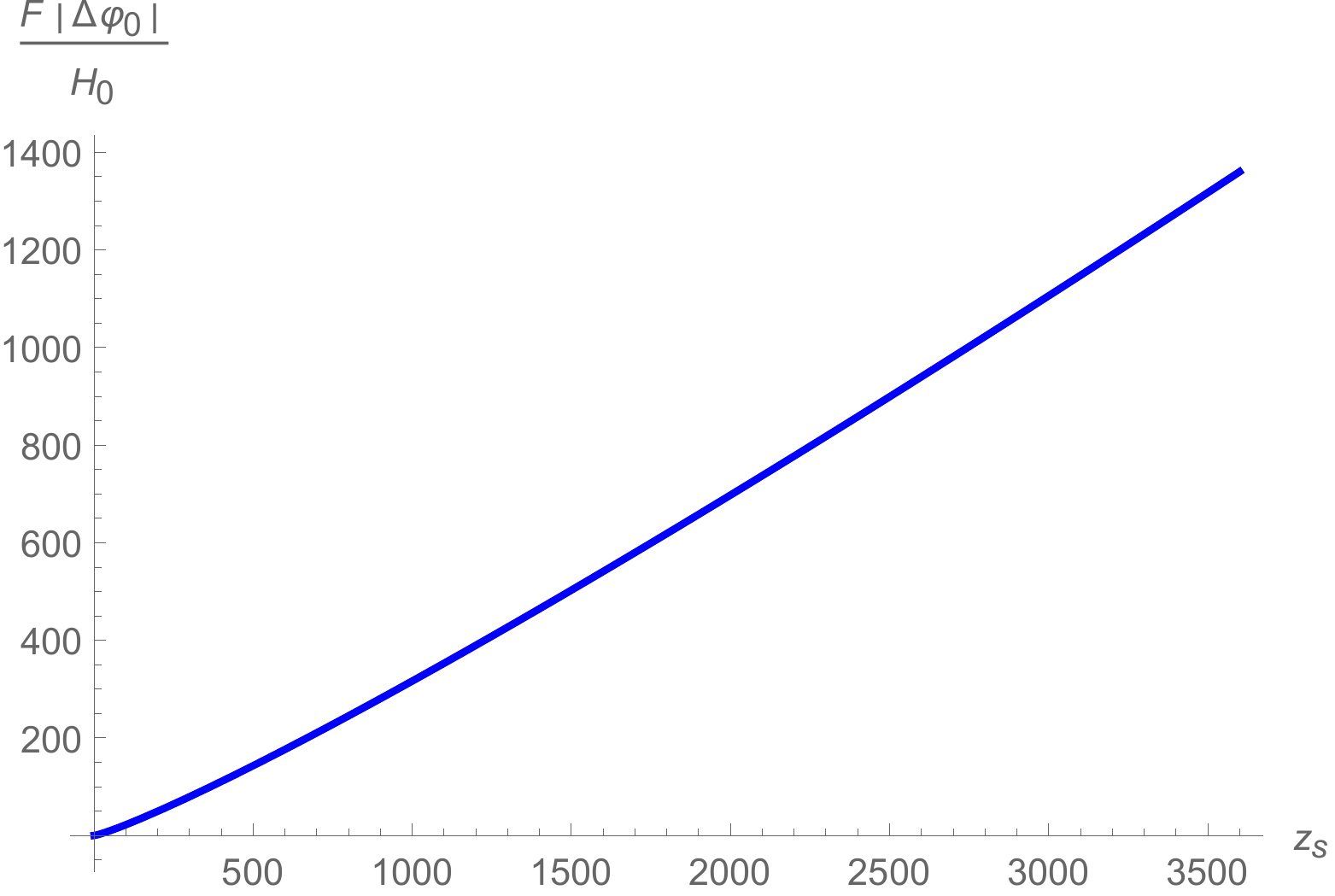}\hspace{0.3cm}
\includegraphics[width=0.3\textwidth]{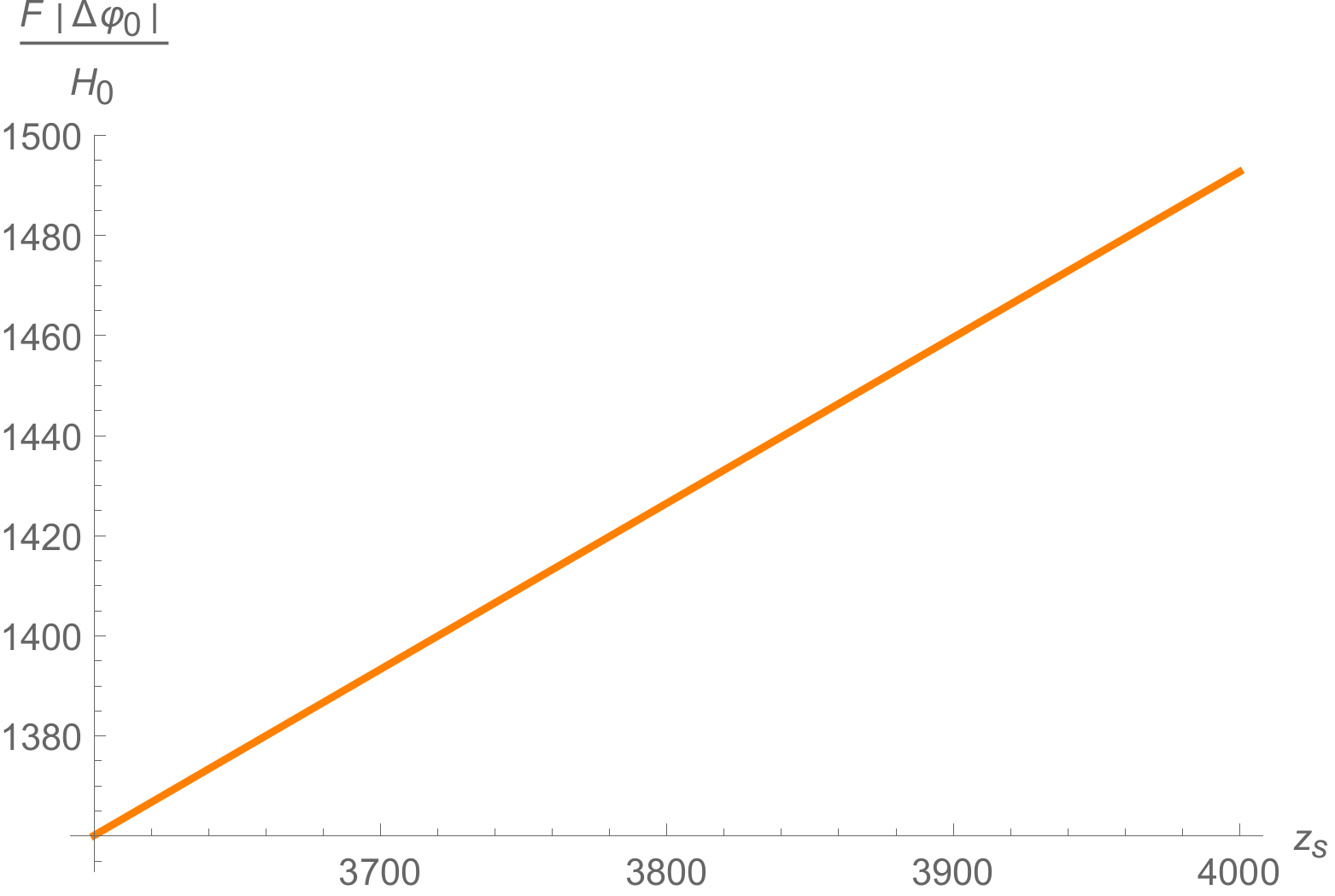}\hspace{0.3cm}
\caption{The rotation angle when $0<z_s<0.4$\;(left), $0.4<z_s<3600$\;(middle), $3600<z_s<4000$\;(right). }\label{fig:GWrotation5}
\end{figure*}

At the end,   we use the Plank\textbf{ 2015 }data \cite{Aghanim:2018eyx}: $\Omega_{\Lambda}=0.6935$, $\Omega_m=0.3065$, $\Omega_R=0$. Thus, the accumulated polarization
 angle  in the whole propagation process are shown in Fig. \ref{fig:GWrotation4} and Fig. \ref{fig:GWrotation5}.

\begin{figure*}[htb]
\centering
\includegraphics[width=0.45\textwidth]{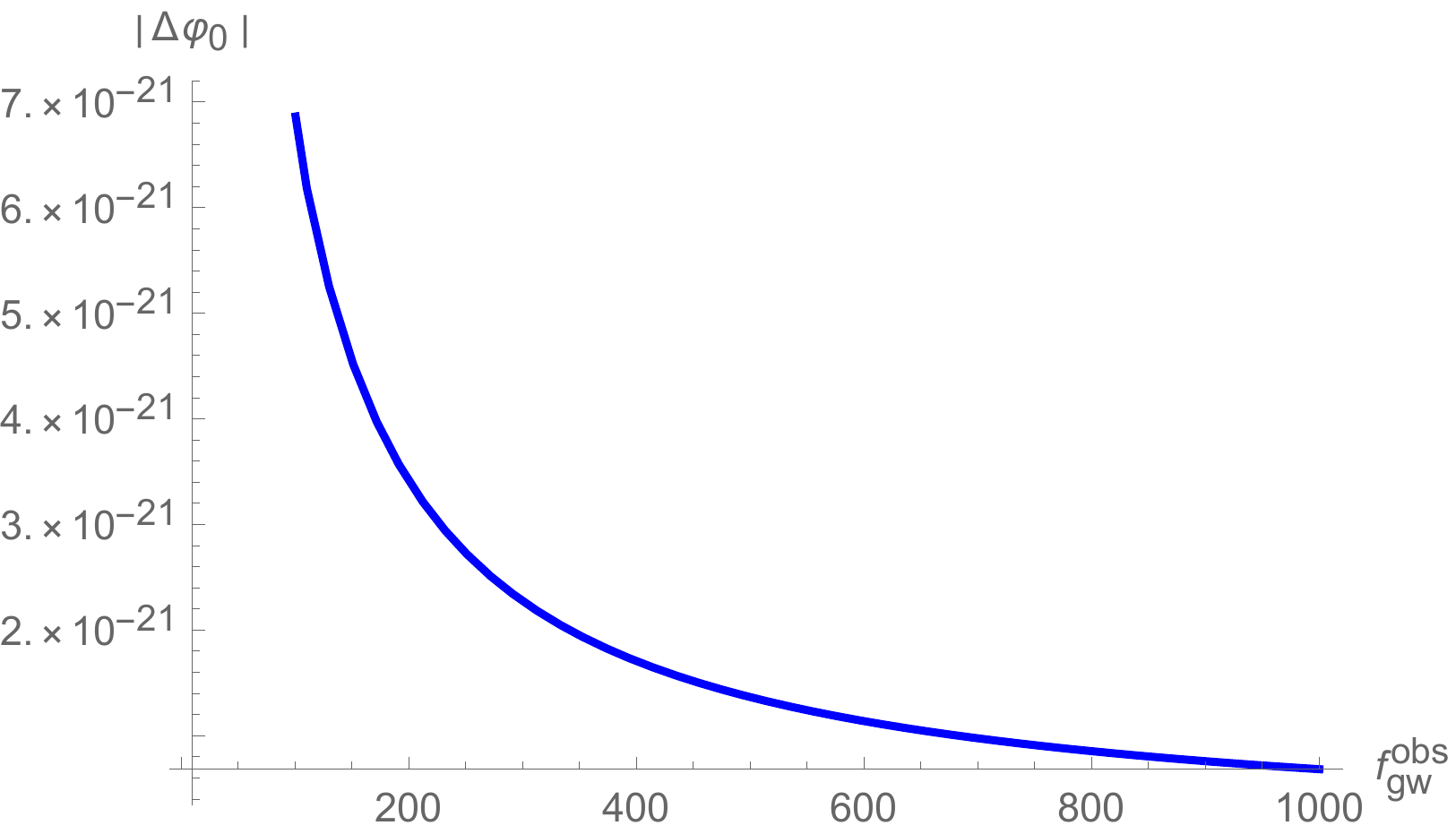}\hspace{0.5cm}
\includegraphics[width=0.45\textwidth]{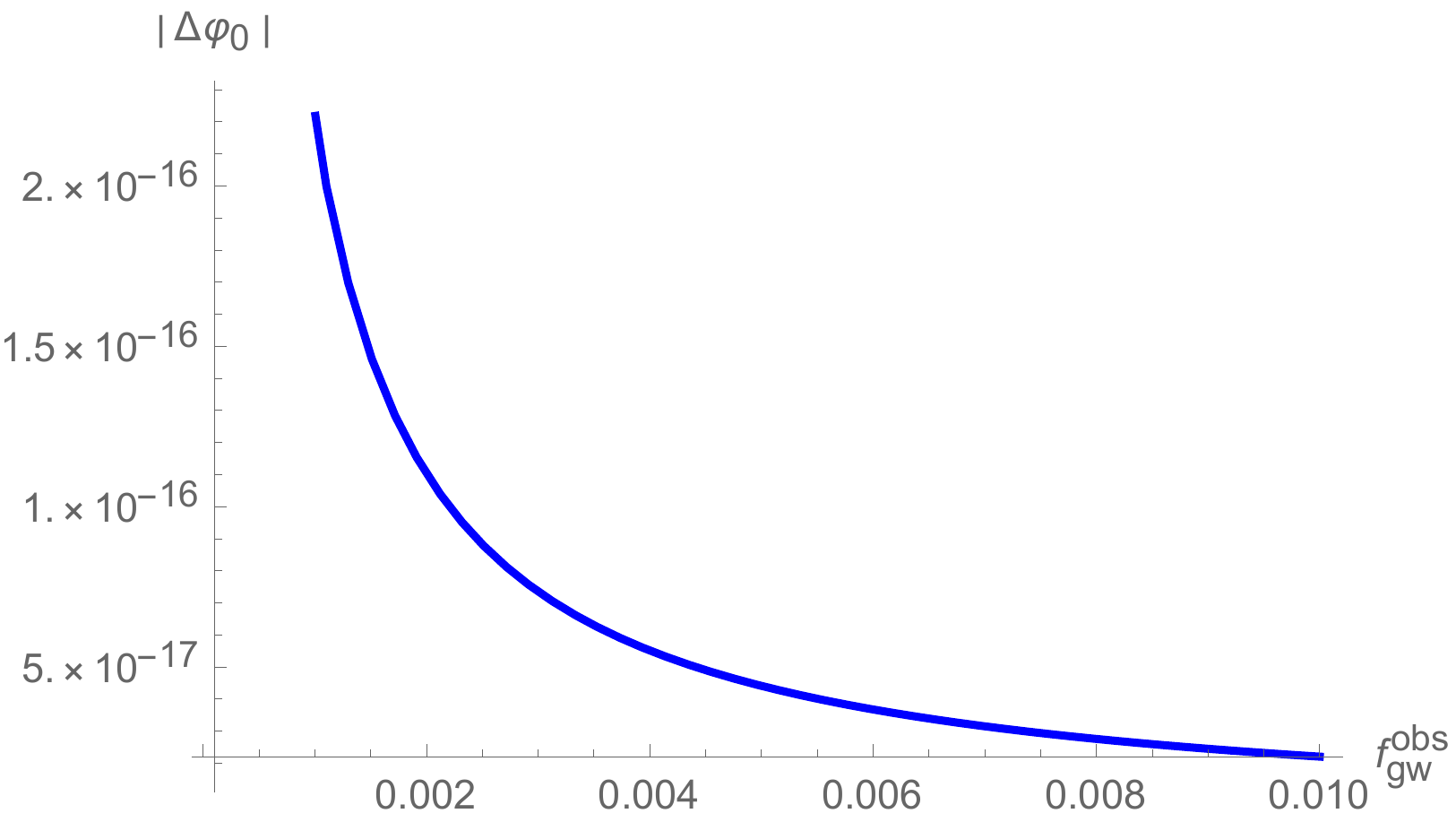}\hspace{0.5cm}
\caption{The rotation angle $\Delta \varphi_0$ is the function of frequency $f^{(obs)}_{gw}$ when we have a given source redshift. The left panel shows the rotation angle  $\Delta \varphi_0$ in  the case that redshift $z_s=60$ and frequency  $10$ Hz~-~$1000$ Hz\;(left). The right panel shows the rotation angle  $\Delta \varphi_0$ in  the case that redshift $z_s=10$ and frequency $10^{-4}$ Hz~-~$10^{-2}$ Hz\;(right). Here, we set $H_0=70~km/s/\mathrm{M}pc$.}\label{fig8}
\end{figure*}

To show the rotation angle more intuitively, let us consider the following two examples:
\begin{itemize}
  \item  The binary system at redshift $z_s=60$. The range of frequency is from $10$ Hz to $1000$ Hz.
        \begin{eqnarray}
        \eta_e &=&\int^{60}_{\infty} \frac{-\mathrm{d}z}{H_0\sqrt{\Omega_{\Lambda}+\Omega_M(1+z)^3+\Omega_R(1+z)^4}},
\end{eqnarray}
\begin{eqnarray}
\left|\Delta \varphi_0\right| &\simeq&\left|-\int^{\eta_m}_{\eta_e}\frac{\sqrt{2}}{\eta^2}\mathrm{d}\eta \; g(v_e)\right|+\left|\int^{\eta_{\Lambda}}_{\eta_m}\frac1{\sqrt{2}\eta^2}\mathrm{d}\eta \; g(v_e)\right|\nb\\
&\simeq& 0.2925 H_0/f^{(obs)}_{gw}.
\end{eqnarray}
We plot $\Delta \varphi_0$ as a function of frequency $f^{(obs)}_{gw}$ in Fig. \ref{fig8} where the source redshift is $z_s=60$.

  \item The binary system  at redshift $z_s=10$. The range of frequency is from $10^{-4}$ Hz to $10^{-2}$ Hz.
 \begin{eqnarray}
        \eta_e &=&\int^{10}_{\infty} \frac{-\mathrm{d}z}{H_0\sqrt{\Omega_{\Lambda}+\Omega_M(1+z)^3+\Omega_R(1+z)^4}},
\end{eqnarray}
\begin{eqnarray}
\left|\Delta \varphi_0\right| &\simeq&\left|-\int^{\eta_m}_{\eta_e}\frac{\sqrt{2}}{\eta^2}\mathrm{d}\eta \; g(v_e)\right|+\left|\int^{\eta_{\Lambda}}_{\eta_m}\frac1{\sqrt{2}\eta^2}\mathrm{d}\eta \; g(v_e)\right|\nb\\
&\simeq& 0.0945461H_0/f^{(obs)}_{gw} .
\end{eqnarray}
We plot $\Delta \varphi_0$ as a function of frequency $f^{(obs)}_{gw}$ in Fig. \ref{fig8} where the source redshift is $z_s=10$.
\end{itemize}

From these quantitative estimates, one can conclude that the value of $|\Delta\varphi|$ is very small, far from the sensitivity of the existing ground-based laser interferometers like LIGO and Virgo. Moreover, the lower the frequency, the more obvious the rotation of the polarization. Hence, it is possible to observe this effect for future GW detectors in extremely low frequency.
\section{Conclusion and Discussion}
\label{sec:conclusion}

In this paper,  we have carried out the detailed analysis on the polarizations  of a GW emitted by an astrophysical source at a cosmic scale distance   and propagating through  the flat FLRW background to the Earth.
In such a curved background,  GWs usually consist of two different parts, the direct and tail. To define the polarization of such a GW, we
have first studied the null geodesic deviations and written them in terms of the Weyl and Ricci scalars, $\Psi_A$ and $\Phi_{AB}$, the projections of the Weyl tensor and Ricci tensor
onto a null tetrad, as the latter have direct physical meaning, which enable us to define the polarization of a GW, given, respectively, by Eqs.(\ref{varphi0})
and (\ref{varphi4}), in terms of the real and imaginary parts of the Weyl scalars, $\Psi_0$ and $\Psi_4$. Since the two spatial unit vectors $e_2$ and $e_3$, which span the polarization plane of the
GW, are parallel-transported along the null geodesics, the change of the polarization angle  is gauge-invariant, that is, if the polarization angle changes in one coordinate system, it will change
in any of them. After detailed analyses, we have shown that effects of the expanding universe on the  ``+" and ``$\times$" modes are different, and usually depend on their waveforms. As a result,
the polarization angle gets changed when the wave propagates through our Universe. In particular, we have found that in different epochs of the universe, the effects are also different.
For a GW emitted by a binary system, we have found explicitly  the relation between the change of the polarization angle $|\Delta \varphi|$
 and the redshift  $z_s$ of the source in different epochs. In particular, in the standard  $\Lambda$CDM model, we have shown  that the order of $|\Delta \varphi|{\eta_0 F}$
  is typically $O(10^{-3})$ to $O(10^3)$, depending on the values of $z_s$, where $\eta_0$ is the (comoving)  time of the current universe, and
  $F\equiv\Big(\frac{5}{256}\frac{1}{\tau_{obs}}\Big)^{3/8}\left(G_NM_c\right)^{-5/8}$ with $\tau_{obs}$ and  $M_c$ being, respectively, the time
  to coalescence in the observer's frame and the chirp mass of the binary system. Particularly, for typical sources of LIGO/Virgo, we find the typical value for $|\Delta \varphi|$ is $10^{-21}$ as shown in Fig. \eqref{fig8}, which is very small and is far from being detected by the current detectors. However, for lower frequency, the observation effect increases linearly. Hence, it is possible for future GW detectors sensitive in extremely low frequency.

It should be noted that there are two timescales: One is related to the propagation of GWs, the other refers to the period of detections of GWs.  The former  can be comparable with the age of the universe for astrophysical GW sources with high redshift, while the latter is of the order of seconds for the current ground-based detectors. Therefore, the changes of polarizations of GWs cannot be directly related to the current detections of GWs.  However, the effects of the background on the propagations and polarizations of such GWs  cannot be ignored as shown in section \ref{3}, especially for sources with low frequency and high redshift.

It would be very interesting to find the observational signature of the rotations of the polarization. However, since the changes are due to the curvature of the background,
  they become significant only over  cosmic scales. Therefore, to detect such effects, combinations of cosmic distant objects with  the ground- and/or space-based detectors might be needed. Interestingly, analysis about the amplitude and phase of GWs traveling a long cosmological distance has been discussed a lot in \cite{Hou:2019jhu}.

  In addition, in this paper we assumed that the GWs are propagating on the  homogeneous and isotropic background. This implies that the wavelengths of such GWs are much shorter than the scale,
  over which the changes of the universe including its perturbations are negligible. This automatically ignores the gravitational lensing of such GWs by massive objects, such as galaxies and supermassive
  black holes. It would be very interesting and important to study such effects, specially their observational evidences.


 \section*{Acknowledgements}

We would like very much to express our gratitude to Prof. Shinji Mukohyama for his long time collaboration on the subjects, and valuable comments and suggestions, which lead us  to sharp
the content considerably, and make the current version more readable. We are very grateful to the anonymous referee for  her/his valuable comments and suggestions.
This work  was partially supported by  the National Natural Science Foundation of China with the Grants Nos. 11975116, and 11975203,
  and Jiangxi Science Foundation for Distinguished Young Scientists under the grant number 20192BCB23007.

\appendix

\section{The Weyl, Ricci Scalars and Gauge Invariance}
\label{appda}
\renewcommand{\theequation}{A.\arabic{equation}}

 \begin{widetext}

In this appendix, we present some detailed computations of   the Weyl and Ricci scalars  in terms of the  null tetrad. In particular, we find,
\bqn
\lb{A1}
\Psi_0 &\equiv& - C_{\mu\nu\lambda\delta}l^{\mu}m^{\nu}l^{\lambda}m^{\delta} \nb\\
&=&  {-}\frac{1}{8 A^2 a^4} \big[-2 h_{+,\eta z} + h_{+,zz}+   h_{+,\eta \eta}-i(-2 h_{{\times},\eta z} + h_{{\times},zz}+ h_{{\times},\eta \eta})\big]\epsilon  + O(\epsilon^2),   \nb\\
 \Psi_1 &\equiv& -C_{\mu\nu\lambda\delta} l^\mu n^\nu l^\lambda m^\delta \nb\\
 &=& {-}\frac{\sqrt{2}}{16a^{3} A} \big[ h_{+,\eta x}- h_{+,xz}+i( h_{+,yz}- h_{+,\eta y})
  -i(h_{{\times},\eta x}- h_{{\times},xz}+i( h_{{\times},yz}- h_{{\times},\eta y}))\big]\epsilon  + O(\epsilon^2), \nb\\
\Psi_2 &\equiv& -\frac{1}{2}C_{\mu\nu\lambda\delta}\big(l^\mu n^\nu l^\lambda n^\delta - l^\mu n^\nu m^\lambda \bar{m}^\delta \big)\nb\\
&=& {-}\frac{1}{12a^2} \big( h_{+,xx}- h_{+,yy} -2h_{{\times},xy}\big)\epsilon  +O(\epsilon^2),  \nb\\
\Psi_3 &\equiv& {-} C_{\mu\nu\lambda\delta} l^\mu n^\nu n^\lambda \bar{m}^\delta \nb\\
&=&  {-}\frac{\sqrt{2}A}{8 a} \big[ h_{+,\eta x} +  h_{+,xz}+i( h_{+,yz}+
 h_{+,\eta y}) +i(h_{{\times},\eta x} +  h_{{\times},xz}+i( h_{{\times},yz}+
 h_{{\times},\eta y}))\big]\epsilon  + O(\epsilon^2),  \nb\\
 \Psi_4 &\equiv& -C_{\mu\nu\lambda\delta}n^{\mu}\bar{m}^{\nu}n^{\lambda}\bar{m}^{\delta} \nb\\
&=&  {-}\frac{A^2}{2} \big[2 h_{+,\eta z}+ h_{+,zz}+ h_{+,\eta \eta}\nb  +i(2 h_{{\times},\eta z}+ h_{{\times},zz}+ h_{{\times},\eta \eta}) \big]\epsilon  + O(\epsilon^2),\\
\eqn
and
 \bqn
\lb{A2}
\Phi_{00} &\equiv&  {-}\frac{1}{2}S_{\mu\nu}l^{\mu}l^{\nu}  \nb\\
&=&  \frac{1}{4 A^2 a^6}\big(2 {a'}^2-a  a'' \big) +O\left(\epsilon^2\right), \nb\\
 \Phi_{11} &\equiv&  {-}\frac{1}{4} S_{\mu\nu}(l^{\mu}n^{\nu}+m^{\mu}\bar{m}^{\nu})\nb\\
&=& \frac{1}{4a^4}\big(2{a'}^2-a  a'' \big)   {-} \frac{1}{8a^2} \big(h_{+,yy} - h_{+,xx} + 2 h_{{\times},yx}\big)\epsilon + O(\epsilon^2),  \nb\\
\Phi_{22} &\equiv& {-} \frac{1}{2} S_{\mu\nu}n^{\mu}n^{\nu}\nb \\
&=& \frac{A^2}{a^2}\big(2{a'}^2 -a  a'' \big)+ O(\epsilon^2),  \nb\\
\Phi_{01} &\equiv&  {-}\frac{1}{2} S_{\mu\nu}l^{\mu}m^{\nu} \nb \\
&=& {-} \frac{\sqrt{2}}{16 a^3 A}\big[ h_{+,xz}- h_{+,\eta x}+i( h_{+,\eta y}-
 h_{+,yz}) -i\left( h_{{\times},xz}- h_{{\times},\eta x}+i( h_{{\times},\eta y}-
 h_{{\times},yz})\right)\big] \epsilon + O(\epsilon^2),   \nb\\
 \Phi_{02} &\equiv&  {-}\frac{1}{2}S_{\mu\nu}m^{\mu}m^{\nu}     \nb  \\
&=&   {-}\frac{1}{4 a^3}\big[a( h_{+,zz} -
 h_{+,\eta\eta}) -2 a '  h_{+,\eta} -i\left(a( h_{{\times},zz} -
 h_{{\times},\eta\eta}) -2 a '  h_{{\times},\eta}\right)\big]\epsilon  +O(\epsilon^2),  \nb\\
\Phi_{12} &\equiv&  {-}\frac{1}{2} S_{\mu\nu}n^{\mu}n^{\nu}  \nb\\
&=&   {-}\frac{\sqrt{2} A}{8 a}\big[- h_{+,xz}- h_{+,\eta x} + i( h_{+,\eta y} + h_{+,yz})
 -i\left(- h_{{\times},xz}- h_{{\times},\eta x} + i( h_{{\times},\eta y} + h_{{\times},yz})\right)\big]\epsilon  + O(\epsilon^2),   \nb\\
\Phi_{\Lambda} &\equiv&  {+}\frac{R}{24}\nb \\
&=&  \frac{ a'' }{4a^3}  {-}\frac{1}{24 a^2}\big( h_{+,yy} - h_{+,xx} + 2h_{{\times},xy}\big)\epsilon  + O(\epsilon^2).
 \eqn

When $\mid z \mid \gg x$ and  $\mid z \mid \gg y$, we find,
\bqn
\lb{A3}
\mid \vec{x} \mid &=&z+\frac{x^2+y^2}{2z}+O\left(\frac{1}{z^3}\right), \quad
v = {\frac1{\sqrt{2}}}(\eta - \mid \vec{x} \mid) =  {\frac1{\sqrt{2}}}\Bigg(\eta - z-\frac{x^2+y^2}{2z}+O\left(\frac{1}{z^3}\right)\Bigg), \nb\\
\cal{R}&=&a(\eta)\mid \vec{x} \mid =a \left(z+\frac{x^2+y^2}{2z}\right) +O\left(\frac{1}{z^3}\right).
\eqn 
Then from \eqref{2.2} we obtain
\bqn
\lb{A4}
h_{\sigma,\eta \eta}&=&\frac{1}{\cal{R}} \Bigg[\frac1{2}{\calpsvv} -\sqrt{2}{\calh} {\calpsv}  +\Big({\calh}^2 -{\calhp} \Big) {\calps}\Bigg] + \frac{1}{a^n}\Bigg[{\frac1{2}\caltsuu} + \frac1{2}{\caltsvv} - \sqrt{2}n{\calh}({\caltsu}+{\caltsv})\nb\\
&&~~~~~~ ~~ +n(n{\calh}^2-{\calhp}){\calts}\Bigg], \nb \\
h_{\sigma,\eta x}&=&\frac{1}{\calr z}\Bigg( -\frac1{2}\calpsvv +\frac1{\sqrt{2}}\calh \calpsv \Bigg)x + \frac{1}{a^n}\Bigg[{\frac1{2}\caltsuu}-\frac1{2}{\caltsvv}
 -\frac1{\sqrt{2}}n{\calh}({\caltsu}-{\caltsv})\Bigg]\frac{x}{z}  +O\left(\frac{1}{z^3}\right),  \nb \\
h_{\sigma,\eta y}&=&\frac{1}{\calr z}\Bigg(-\frac1{2}\calpsvv +\frac1{\sqrt{2}}\calh \calpsv \Bigg)y + \frac{1}{a^n}\Bigg[{\frac1{2}\caltsuu}-\frac1{2}{\caltsvv}
 -\frac1{\sqrt{2}}n{\calh}({\caltsu}-{\caltsv})\Bigg]\frac{y}{z}  +O\left(\frac{1}{z^3}\right),  \nb \\
h_{\sigma,\eta z}&=&-\frac1{2}\frac{\calpsvv}{\calr} +\frac1{\sqrt{2}}\left(\calh-\frac{a}{\calr} \right)\frac{\calpsv}{\calr} +a \calh \frac{\calps}{\calr^{2}}  + \frac{1}{a^n}\Bigg[\frac1{2}{\caltsuu}-\frac1{2}{\caltsvv} -\frac1{\sqrt{2}}n{\calh}({\caltsu}-{\caltsv})\Bigg]\nb\\
&&\times \left(1-\frac{x^2+y^2}{2z^2}\right)+O\left(\frac{1}{z^3}\right),  \nb\\
h_{\sigma,xx}&=&-\frac1{\sqrt{2}}\frac{\calpsv}{\calr z}  + \frac{1}{a^n}\Bigg[\frac1{2}({\caltsuu}+{\caltsvv})\frac{x^2}{z^2} +\frac1{\sqrt{2}}({\caltsu}-{\caltsv})\frac{1}{z}\Bigg] +O\left(\frac{1}{z^3}\right), \nb \\
h_{\sigma,yy}&=&-\frac1{\sqrt{2}}\frac{\calpsv}{\calr z}  + \frac{1}{a^n}\Bigg[\frac1{2}({\caltsuu}+{\caltsvv})\frac{y^2}{z^2} +\frac1{\sqrt{2}}({\caltsu}-{\caltsv})\frac{1}{z}\Bigg] +O\left(\frac{1}{z^3}\right), \nb \\
h_{\sigma,zz}&=&\frac1{2}\frac{\calpsvv}{\calr}+\frac{\sqrt{2}a \calpsv}{\calr^2} + \frac{1}{a^n}\left(\frac1{2}{\caltsuu}+\frac1{2}{\caltsvv}\right)\left(1-\frac{x^2+y^2}{z^2}\right)  +O\left(\frac{1}{z^3}\right), \nb\\
h_{\sigma,xz}&=&\frac1{2}\frac{\calpsvv}{\calr z}x + \frac{1}{a^n}\Bigg[\frac1{2}({\caltsuu}+{\caltsvv})\frac{x}{z} +\frac1{\sqrt{2}}({\caltsv}-{\caltsu})\frac{x}{z^2}\Bigg] +O\left(\frac{1}{z^3}\right),  \nb\\
h_{\sigma,yz}&=&\frac1{2}\frac{\calpsvv}{\calr z}y  + \frac{1}{a^n}\Bigg[\frac1{2}({\caltsuu}+{\caltsvv})\frac{y}{z} +\frac1{\sqrt{2}}({\caltsv}-{\caltsu})\frac{y}{z^2}\Bigg] +O\left(\frac{1}{z^3}\right).
\eqn
Inserting them into Eqs.(\ref{A1}) and \eqref{A2}, we  get Eqs.(\ref{RePsi0}), (\ref{ImPsi0}) and
\bqn
\lb{RePhi02}
{\mbox{Re}}\left(\Phi_{02}\right) &=& {-}\frac{1}{4 a^2}\Bigg\{\frac{1}{\calr}\Big[({\calh}^{2}+{\calhp}){\calp}\Big]+\frac{1}{a^n}\Bigg[\frac12\left({\caltuu}+{\caltvv}\right)\left(-\frac{x^2+y^2}{z^2}\right) +\sqrt{2}n{\calh}({\caltu}+{\caltv})~~~~\nb\\
&&~~~~~~~~~~~~~~ -n(n{\calh}^2-{\calhp}){\calt}-\sqrt{2}{\calh}({\caltu}+{\caltv})+2n{\calh}^2{\calt}\Bigg]\Bigg\}\epsilon + {\cal{O}}\left(\epsilon^2\right),
\eqn
\bqn
\lb{ImPhi02}
 {\mbox{Im}}\left(\Phi_{02}\right)  &=& \frac{1}{4 a^2}\Bigg\{\frac{1}{\calr}\Big[({\calh}^{2}+{\calhp}){\calpt}\Big]+\frac{1}{a^n}\Bigg[\frac12\left({\calttuu}+{\calttvv}\right)\left(-\frac{x^2+y^2}{z^2}\right) +\sqrt{2}n{\calh}({\calttu}+{\calttv})~~~~\nb\\
&&~~~~~~~~~~~~ -n(n{\calh}^2-{\calhp}){\caltt}-\sqrt{2}{\calh}({\calttu}+{\calttv})+2n{\calh}^2{\caltt}\Bigg]\Bigg\}\epsilon  + {\cal{O}}\left(\epsilon^2\right).
\eqn

 When it comes to gauge invariance, there are two kinds of gauge transformations:
\begin{itemize}
  \item Gauge transformations of the first kind.
  From Eqs.(\ref{A1}-\ref{A2}), it is clear that the physical quantities such as $\Psi_0$  and $ \Psi_4$ are invariant under the  change of coordinate system
  because the equation of NP formalism just involve scalar functions. These are  the so-called {\it{gauge transformations of the first kind}} as defined in \cite{Sachs:1964zza}.
  \item  Gauge transformations of the second kind.
{\it{The gauge transformations of the second kind}} have been discussed in \cite{Stewart:1974uz,Miranda:2014vaa}, which are changes of the identification map,
 $x^{\mu}_{\mbox{\scriptsize{new}}}=x^{\mu}_{\mbox{\scriptsize{old}}}+\epsilon\xi^{\mu}$, where  $\xi^{\mu}$  is an arbitrary vector field.  Then,  for the liner-order perturbation of a NP quantity $L$, we have $L^{(1)\mbox{\scriptsize{new}}}=L^{(1)\mbox{\scriptsize{old}}}-\epsilon L^{(1)\mbox{\scriptsize{old}}}_{,\mu}\xi^{\mu}$.
 As shown  in \cite{Stewart:1974uz,Miranda:2014vaa}, there are three classes of infinitesimal changes in the tetrad:
\bqn
\begin{aligned}
&\text{(i)}~~~l^{\mu}\rightarrow l^{\mu}, m^{\mu}\rightarrow m^{\mu}+\epsilon al^{\mu}, n^{\mu}\rightarrow n^{\mu}+\epsilon\bar{a}m^{\mu}+\epsilon a\bar{m}^{\mu}+\epsilon^2a\bar{a}l^{\mu};\\
&\text{(ii)}~~~n^{\mu}\rightarrow n^{\mu}, m^{\mu}\rightarrow m^{\mu}+\epsilon bn^{\mu}, l^{\mu}\rightarrow l^{\mu}+\epsilon\bar{b}m^{\mu}+\epsilon b\bar{m}^{\mu}+\epsilon^2b\bar{b}n^{\mu};\\
&\text{(iii)}~~~l^{\mu}\rightarrow Al^{\mu},n^{\mu}\rightarrow A^{-1}n^{\mu}, m^{\mu}\rightarrow e^{i\theta}m^{\mu},
\end{aligned}
\eqn
where $a$ and $b$ are complex functions and $A$ and $\theta$ are real functions. The components of the Weyl tensor, $ \Psi_1, \Psi_2, \Psi_3$ and $\Psi_4$, can be expressed in terms of twelve spin coefficients \cite{Griffiths76a,Griffiths:1991zp,NP62}. Let us  take the class (i) as an example,
then we find that the spin coefficients \cite{Griffiths76a,Griffiths:1991zp,NP62} transform as
\begin{align}
&\kappa\rightarrow\kappa,\;\sigma\rightarrow\sigma+\epsilon a\kappa,\;\rho\rightarrow\rho+\epsilon\bar{a}\kappa,\;\varepsilon\rightarrow\varepsilon+\epsilon\bar{a}\kappa,\notag\\
&\tau\rightarrow\tau+\epsilon a\rho+\epsilon\bar{a}\sigma+\epsilon^2a\bar{a}\kappa,\;\pi\rightarrow\pi+\epsilon D\bar{a}+2\epsilon\bar{a}\varepsilon+\epsilon^2(\bar{a})^{2}\kappa,\notag\\
&\alpha\rightarrow\alpha+\epsilon\bar{a}(\rho+\varepsilon)+\epsilon^2(\bar{a})^{2}\kappa,\;\beta\rightarrow\beta+\epsilon a\varepsilon+\epsilon\bar{a}\sigma+\epsilon^2a\bar{a}\kappa,\notag\\
&\gamma\rightarrow\gamma+\epsilon a\alpha+\epsilon\bar{a}(\beta+\tau)+\epsilon^2a\bar{a}(\rho+\varepsilon)+\epsilon^3(\bar{a})^{2}\sigma+\epsilon^3a(\bar{a})^{2}\kappa,\notag\\
&\lambda\rightarrow\lambda+\epsilon\bar{a}(2\alpha+\pi)+\epsilon^2(\bar{a})^{2}(\rho+2\varepsilon)+\epsilon^3(\bar{a})^{3}\kappa+\epsilon\bar{\delta}\bar{a}+\epsilon^2\bar{a}D\bar{a},\notag\\
&\mu\rightarrow\mu+\epsilon
a\pi+2\epsilon\bar{a}\beta+2\epsilon^2a\bar{a}\varepsilon+\epsilon^2(\bar{a})^{2}\sigma+\epsilon^3a(\bar{a})^{2}\kappa+\epsilon\delta\bar{a}+\epsilon^2aD\bar{a},\notag\\
&\nu\rightarrow\nu+\epsilon a\lambda+\epsilon\bar{a}(\mu+2\gamma)+\epsilon^2(\bar{a})^{2}(\tau+2\beta)+\epsilon^3(\bar{a})^{3}\sigma+\epsilon^2a\bar{a}(\pi+2\alpha)\notag\\
&\;\;\:\quad+\epsilon^3a(\bar{a})^{2}(\rho+2\varepsilon)+\epsilon^4a(\bar{a})^{3}\kappa+\epsilon\big(\Delta+\epsilon\bar{a}\delta+\epsilon a\bar{\delta}+\epsilon^2a\bar{a}D\big)\bar{a}.
\label{spin-I}
\end{align}
Thus, $\Psi_0,  \Psi_1, \Psi_2, \Psi_3$ and $\Psi_4$ under the transformation transfer as
\bqn
\begin{aligned}
&\Psi_{0}\rightarrow\Psi_{0},\;\Psi_{1}\rightarrow\Psi_{1}+\epsilon\bar{a}\Psi_{0},\;\Psi_{2}\rightarrow\Psi_{2}+2\epsilon\bar{a}\Psi_{1}+\epsilon^2(\bar{a})^{2}\Psi_{0},\\
&\Psi_{3}\rightarrow\Psi_{3}+3\epsilon\bar{a}\Psi_{2}+3\epsilon^2(\bar{a})^{2}\Psi_{1}+\epsilon^3(\bar{a})^{3}\Psi_{0},\\
&\Psi_{4}\rightarrow\Psi_{4}+4\epsilon\bar{a}\Psi_{3}+6\epsilon^2(\bar{a})^{2}\Psi_{2}+4\epsilon^3(\bar{a})^{3}\Psi_{1}+\epsilon^4(\bar{a})^{4}\Psi_{0}.\\
\end{aligned}
\label{Weyl-Maxwell-I}
\eqn
 Meanwhile, we find that $\Psi_1, \Psi_2, \Psi_3$ in Eqs.\eqref{A1}  vanish for the metric of Eq.\eqref{gwmetric} in the propagation region far from the source, $\left|\vec{x}\right| \gg D$,
 so the effect to the linear order of $\epsilon$ of $ \Psi_0$  and $ \Psi_4$ is
  \bq
\Psi_{0}^{\scriptscriptstyle{(1)}}\rightarrow\Psi_{0}^{\scriptscriptstyle{(1)}},\quad \Psi_{4}^{\scriptscriptstyle{(1)}}\rightarrow\Psi_{4}^{\scriptscriptstyle{(1)}}.
\label{Weyl-Maxwell-I2}
\eq
Other classes can be studied by following similar considerations, and we find that  the results are the same, so we do not show the detail here.
\end{itemize}
 In summary, $\Psi_0$  and $ \Psi_4$ are invariant under any kind of the gauge transformations.
 \end{widetext}

 \section{Parallel-transported Polarization Angle}
 \label{appdb1}
\renewcommand{\theequation}{B.\arabic{equation}}\setcounter{equation}{0}

It was shown in \cite{Wang91} that the above definition of the polarization angle is in general not parallel-transported along the wave paths. This makes it impossible to compare the polarization of a  gravitational wave at two different points along its propagation path. One way to overcome this difficulty is to find a parallel-transported basis carried by the wave. Once we have this basis, polarization angle defined based on this basis should be parallel-transported. As suggested in \cite{Wang91}, a parallel-transported basis can be found by rotating the coordinate with a proper angle in the $(e_2^{\mu},e_3^{\mu})$ plane.
Similar considerations can be generalized to the current case. In particular, for GWs along the null geodesics defined by  $n^{\mu}$, we let $\varphi_0^{(0)}$ be the angle of rotating, and $\lambda_{(2)}^{\mu}$ and $\lambda_{(3)}^{\mu}$ be the new basis. Then, we find
\bqn
\lambda_{(2)}^{\mu}{}_{;\nu}l^{\nu}&=&\left(\frac12\sinh{W}V_{,\nu}-\varphi_{0,\nu}^{(0)}\right)l^{\nu}\lambda_{(3)}^{\mu},\\
\lambda_{(3)}^{\mu}{}_{;\nu}l^{\nu}&=&-\left(\frac12\sinh{W}V_{,\nu}-\varphi_{0,\nu}^{(0)}\right)l^{\nu}\lambda_{(2)}^{\mu},
\eqn
which implies
\bq
\varphi_{0,\nu}^{(0)}=\frac12\sinh{W}V_{,\nu}= \epsilon^2 h_{\times}h_{+,\nu}+\mathcal{O}(\epsilon^3).
\eq
Then, the angle
\bq
\psi_0\equiv \hat\varphi_0-\varphi_{0}^{(0)},
\eq
determines the polarization direction of the $\Psi_0$ wave relative to the $(\lambda_{(2)}^{\mu},\lambda_{(3)}^{\mu})$ basis.
Similarly, for the $\Psi_4$ wave, the related angle is given by
\bq
\varphi_{4,\nu}^{(0)}= \epsilon^2 h_{\times}h_{+,\nu}+\mathcal{O}(\epsilon^3),
\eq
and $\psi_4\equiv \hat\varphi_4-\varphi_{4}^{(0)}$  defines the angle between the polarization direction of the wave and the $\lambda_{(2)}^{\mu}$ axis.
It should be noted that both $\varphi_{0}^{(0)}$ and $\varphi_{4}^{(0)}$ are $\epsilon^2$-term, which implies that  there is no linear-order corrections.

\section{Timelike Geodesics Deviations and Polarization of a Gravitational Wave }\label{appdb}
\renewcommand{\theequation}{C.\arabic{equation}}\setcounter{equation}{0}

In this appendix, let us consider timelike geodesic deviations, as they are directly related to observations. To our purpose, in the following we consider timelike geodesic congruences formed by $t^{\mu}$, from Eqs.(\ref{B.1}) we find
\bq
\lb{B.1b}
t^{\mu}{}_{;\nu}t^{\nu}=-\frac1{\sqrt{2}}\left(\frac{AB_{,\eta}}{B}-\frac{BA_{,\eta}}{A}\right)s^{\mu}.
\eq
Therefore, choosing $A$ and $B$ so that
\bq
\label{ABeq}
\frac{AB_{,\eta}}{B}-\frac{BA_{,\eta}}{A}=0,
\eq
we find $t^{\mu}{}_{;\nu}t^{\nu}=0$, that is,  $t^{\mu}$ defines a timelike and affinely parametrized geodesic congruence. Recall that  $2AB=a^{-2}$. Thus, the general solution of the above
equation is
\bqn
\lb{B.1c}
A=\frac{-A_0\pm\sqrt{A_0^2+2a^2}}{-2a^2},\quad
B=\frac1{2Aa^2},
\eqn
where $A_0$ is an integration constant. Without loss of the generality, we can always choose  $A_0=0$, so that
\bq
\lb{B.1d}
A=B=\pm \frac{1}{\sqrt{2}\; a}.
\eq
 In what follows, we shall adopt the ``+" sign.

On the other hand, the four spatial unity vectors we constructed
in section \ref{2}, form an orthogonal base. Thus,  the  component $\Psi_0$ ($\Psi_4$)
represents the GW moving along the positive (negative) $s^{\mu}$-direction,
as shown in Fig. \ref{fig:Ptz}.

Let $\eta^{\mu}$ be the geodesic deviation between two neighbor geodesics and $\eta^{\mu} t_{\mu} = 0$. Then,
the timelike geodesic deviation is given by,
 \bqn
\lb{GeoDev1}
\frac{d^2\eta^{\mu}}{d\lambda^2}&=&- R^{\mu}_{\nu\lambda\beta} t^{\nu}\eta^{\lambda}t^{\beta} \nb\\
&=&  \frac12\Big\{-\left(\Phi_{00}+2\Phi_{11}+\Phi_{22}-6\Phi_{\Lambda}\right) e^{\mu\nu}_{o}  \nb\\
&&+ \Big[\left(\Psi_2+\bar{\Psi}_2\right)+2\left(\Phi_{\Lambda}-\Phi_{11}\right)\Big]\left(2s^{\mu}s^{\nu} - e^{\mu\nu}_{0}\right)\nb\\
&&  {+}  {2}{\mbox{Re}}\Big[\left(\Psi_1+  \Psi_3\right) - \left(\Phi_{01} - \Phi_{12}\right)\Big]e^{\mu\nu}_{\times  2s} \nb\\
&&   {+}2 {\mbox{Im}}\Big[\left(\Psi_1-  \Psi_3\right) - \left(\Phi_{01}  -\Phi_{12}\right)\Big]e^{\mu\nu}_{\times  3s}\nb\\
&&  {+}{\mbox{Re}} \left(\Psi_0+\Psi_4 + 2\Phi_{02}\right)\; e^{\mu\nu}_{+} \nb\\
&&  + {\mbox{Im}} \left(\Psi_0- \Psi_4 +2\Phi_{02}\right)\; e^{\mu\nu}_{\times}\Big\} \eta_{\nu},
\eqn
 where
\bqn
\lb{B2.b}
e^{\mu\nu}_{\times  2s} &\equiv& e^{\mu}_2s^{\nu} + e^{\nu}_{2}s^{\mu},\quad
e^{\mu\nu}_{\times  3s}\equiv e^{\mu}_3s^{\nu} + e^{\nu}_{3}s^{\mu},\nb\\
\eqn

 The right hand side of Eq.\eqref{GeoDev1} includes six terms, each of them represents a sort of polarization mode, and has the following physical interpretations \cite{Szekeres65,WThesis}. The first terms is transverse,
 laying on the ($e_2, e_3$)-plane only, and makes the test particles contract or expand uniformly,  depending on sign of the coefficient. It is remarkable to note that the Weyl scalars have no contributions to this term.
 The second term   has the effect of distorting a sphere of particles about the observer into an ellipsoid,
which has $s^{\mu}$ as principal axis, while in the  perpendicular ($e_2, e_3$)-plane it is uniformly contracting or expanding, depending the signs of the coefficient,
$\left(\Psi_2+\bar{\Psi}_2\right)+2\left(\Phi_{\Lambda}-\Phi_{11}\right)$. This is a behavior quite similar to  particles falling in towards
a central attracting body with the inverse square law, so it is often to call this term as the Coulomb part of the field, and  the strength of the Coulomb field of the pure gravitational field
is given by the real part of $\Psi_2$, while the strength of   the matter field is represented by $\left(\Phi_{\Lambda}-\Phi_{11}\right)$.
 The third term represents a force that acts only on the ($e^{\mu}_2, s^{\mu}$)-plane. Recall $s^{\mu}$ represents the direction of the propagation of the two GW components, represented by $\Psi_0$ and $\Psi_4$,
 respectively, so this term is often called  the longitudinal component. Rotating this plane by $45^0$,
 \bqn
 \lb{B.2c}
 s^{\mu} &=& \cos 45^0 {s'}^{\mu} + \sin 45^0 {e'}^{\mu}_2,\nb\\
  e^{\mu}_2 &=& -\sin 45^0 {s'}^{\mu} + \cos 45^0 {e'}^{\mu}_2,
 \eqn
 we find that
 \bq
  \lb{B.2d}
 e^{\mu\nu}_{\times  2s} = -\left({s'}^{\mu}{s'}^{\nu}  - {e'}^{\mu}_2{e'}^{\nu}_2\right).
 \eq
 Therefore, if it is initially a circle in the ($e^{\mu}_2, s^{\mu}$)-plane, this term will make the circle into an ellipse with its major axis along the $45^0$ with respect to the $s^{\mu}$-direction. The strength of this force
 depends on the real part of the combination, $\left(\Psi_1+  \Psi_3\right)  - \left(\Phi_{01} - \Phi_{12}\right)$.

 In contrast to the third term, the fourth term represents a force that acts only on the ($e^{\mu}_3, s^{\mu}$)-plane, and the strength of this term depends on the imaginary  part of the combination,
 $\left(\Psi_1 -  \Psi_3\right)  - \left(\Phi_{01} - \Phi_{12}\right)$. Note the difference between this term and the last one in the pure gravitational part, represented by $\Psi_1$ and $\Psi_3$, respectively. In the third term,
 it is the real part of $\left(\Psi_1 +  \Psi_3\right)$, while in the fourth term, it is the imaginary part of $\left(\Psi_1 -  \Psi_3\right)$. This is different from the matter field, which is always proportional to the combination of
 $ \left(\Phi_{01} - \Phi_{12}\right)$ in both terms.

The fifth term is also purely transverse and deforms a circle laying in the ($e_2, e_3$)-plane into ellipse with its major axis along the $e_2$-direction.
To see the effects of the last term, we can make a similar rotation as that given by Eq.(\ref{B.2c})  but now in the  ($e_2, e_3$)-plane, so we find that
\bq
\lb{B.2e}
 e^{\mu\nu}_{\times} = - {e'}^{\mu\nu}_+.
\eq
Therefore,  the last term is also purely transverse and deforms a circle laying in the ($e_2, e_3$)-plane into ellipse with its major axis along a line that has a $45^0$ angle with respect to the $e_2$-direction.

 To consider the polarization of the   right-moving (along the positive direction of $s^{\mu}$) GW represented by $\Psi_0$, let us  first make the rotation as Eq.(\ref{rotation1}),
\bqn
\lb{rotation2}
e_2^{\mu}  &=& \cos\varphi_0 \hat{e}_2^{\mu} + \sin\varphi_0 \hat{e}_3^{\mu},\nb\\
e_{3}^{\mu}  &=& - \sin\varphi_0 \hat{e}_2^{\mu} + \cos\varphi_0 \hat{e}_3^{\mu},
\eqn
where
\bq
\lb{Varphi0}
\tan 2\varphi_0 \equiv - \frac{{\mbox{Im}}\left(\Psi_0\right)}{{\mbox{Re}}\left(\Psi_0\right)},
\eq	
then we find that the $\Psi_0$ parts in the right-hand side of Eq.(\ref{GeoDev1}) become,

\bqn
\lb{GeoDev1.1}
 {\mbox{Re} \left(\Psi_0\right)}\; e^{\mu\nu}_{+}  +  {\mbox{Im}} \left(\Psi_0\right)\; e^{\mu\nu}_{\times}
= \left(\Psi_0\bar\Psi_0\right)^{1/2} \hat{e}^{\mu\nu}_{+}.~~~~~~
  \eqn
That is, the $\Psi_0$ GW makes a circle in  ($e_2, e_3$)-plane into ellipse with its major axis along the $\hat e_2$-direction. Such a defined angle is
referred as to the polarization angle of the $\Psi_0$ GW \cite{Wang91,WThesis}.

  Note that, to the first-order of $h_\sigma$, the two spatial unit vectors $e_2$ and $e_3$ are  of parallel transport along the time geodesics defined by $t^{\mu}$,
\bq
t^{\nu}\nabla_{\nu}e^{\mu}_2 \simeq 0 +  {\cal{O}}\left(\epsilon^2\right), \quad t^{\nu}\nabla_{\nu}e^{\mu}_3  \simeq 0 +   {\cal{O}}\left(\epsilon^2\right),
\eq
so that the angle $\varphi_0$ defined above has an absolute meaning.

It must be noted that  the matter component $\Phi_{02}$ has  similar effects on the circle laying in the ($e_2, e_3$)-plane, as we can see from the last two terms in Eq.(\ref{GeoDev1}). Therefore,  this
terms also makes the circle into an ellipse, with the same effects precisely as the $\Psi_0$ GW component, and in principle the detectors, such as LIGO and Virgo, cannot distinguish these effects, and will measure
them together. However, from Eqs.(\ref{RePsi0})-(\ref{eq2.16}) and (\ref{RePhi02})-(\ref{ImPhi02}), we find that, relative to $\Psi_0$, the matter component $\Phi_{02}$ is either suppressed by a factor ${\cal{H}}$ or $(x^2+y^2)/z^2$. So, comparing with
$\Psi_0$, the effects of $\Phi_{02}$ are negligible, and can be safely ignored for the current and forthcoming detectors, including the third generation of detectors.

With the above in mind,   for the $\Psi_4$ wave,  if we make a similar rotation as that given by Eq.(\ref{rotation2}) but now with the angle given by,
\bq
\lb{Varphi4}
\tan 2\varphi_4 \equiv  {+} \frac{{\mbox{Im}}\left(\Psi_4\right)}{{\mbox{Re}}\left(\Psi_4\right)},
\eq	

we find that \cite{Wang91,WThesis},

\bqn
\lb{GeoDev1.2}
  {\mbox{Re}} \left(\Psi_4\right)\; e^{\mu\nu}_{+}   -  {\mbox{Im}} \left(\Psi_4\right)\; e^{\mu\nu}_{\times}
  = \left(\Psi_4\bar\Psi_{4}\right)^{1/2} \hat{e}^{\mu\nu}_{+}, ~~~~~~
  \eqn
and such defined angle $\varphi_4$ is referred to as the polarization angle of the $\Psi_4$ wave. Again, in the current case  it has an absolute meaning, as
the frame defined by $e_2$ and $e_3$ are  of parallel transport along the time geodesics.

In summary, the definition of the polarization of a  GW is valid in any given spacetime, independent of its symmetry and the nature of the geodesic deviations, null or timelike.

\section{The expression of ${\cal{P}}_{\sigma}$}\label{appdc}
\setcounter{equation}{0}
\setcounter{subsection}{0}
\renewcommand{\theequation}{D.\arabic{equation}}
\renewcommand{\thesubsection}{D.\arabic{subsection}}

\begin{figure}[htb]
\centering
\includegraphics[scale=0.36]{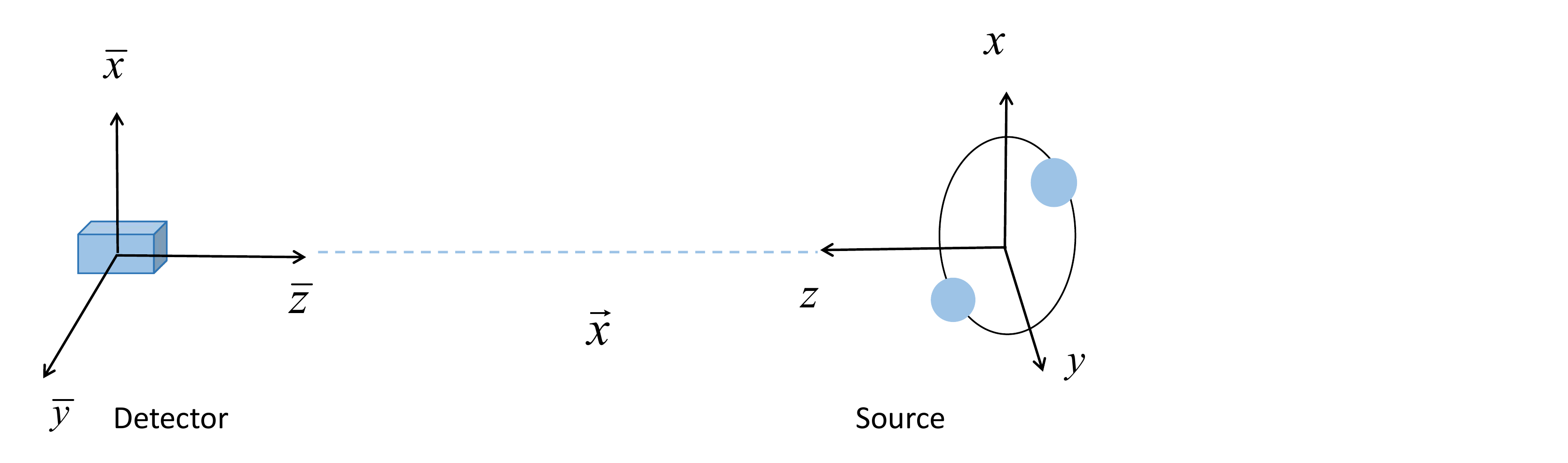}
\caption{Sketch of THE propagation of GWs from a source to the detector. $(x, y, z)$ is the coordinate system of the source frame and $(\bar{x},\bar{y},\bar{z})$ is the coordinate system of observer's frame.} \label{fig:Pds}
\end{figure}

To calculate  ${\cal{P}}_{\sigma}$,  we first note that
\bq
\Pi_{ij}=\Lambda_{ij,kl}\delta T^{kl},
\eq
where $\Lambda_{ij,kl}(\vec{n})=P_{ik}P_{jl}-\frac12 P_{ij}P_{kl}$ and $P_{ij}(\vec{n})=\delta_{ij}-n_in_j$.
Then we get
\bqn
F_{ij}^{(\gamma)} &=& \Lambda_{ij,kl}\int_{\mathbb{R}^3}{\mathrm{d}^3\vec{x}'a^3(v) \delta T^{kl}\left(v, \vec{x}'\right)},\\
&=& \Lambda_{ij,kl}S^{kl},
\eqn
where $S^{ij}=\int_{\mathbb{R}^3}{d^3\vec{x}'a^3(v) \delta T^{ij}\left(v, \vec{x}'\right)}$.
We also define
\bq
M^{ij}\equiv \int_{\mathbb{R}^3}{\mathrm{d}^3\vec{x}'a^3(v) \delta T^{00}\left(v, \vec{x}'\right)x^ix^j}.
\eq
It is not difficult to prove that the following relation holds
\bq
S^{ij}=\frac12\ddot{M}^{ij},
\eq
where a dot denotes the derivative with respect to (w.r.t.) the cosmic time  $t$, with $dt \equiv a(\eta) d\eta$.
Following \cite{Maggiore:1900zz}, with
\bq
\vec{n}=(\sin{\theta}\sin{\phi},\sin{\theta}\cos{\phi},\cos{\theta}),
\eq
 we find
\bqn
\nb D^{(\gamma)}_{+}(\eta;\theta,\phi)&=&\frac{ G_N }{ \left|\vec{x}\right| a(\eta)}\left[\ddot{M}_{11}(\cos^2{\phi}-\sin^2{\phi}\cos^2{\theta})\right.\\
&&\nb \left.+\ddot{M}_{22}(\sin^2{\phi}-\sin^2{\phi}\cos^2{\theta})\right.\\
&&\nb \left.-\ddot{M}_{33}\sin^2{\theta}-\ddot{M}_{12}\sin{2\phi}(1+\cos^2{\theta})\right.\\
&&\left.+\ddot{M}_{13}\sin{\phi}\sin{2\theta}+\ddot{M}_{23}\cos\phi\sin{2\theta}\right],\nb\\
\eqn
and
\bqn
\nb D^{(\gamma)}_{\times}(\eta;\theta,\phi)&=&\frac{ G_N }{ \left|\vec{x}\right| a(\eta)}\left[(\ddot{M}_{11}-\ddot{M}_{22})\sin2\phi\cos\theta\right.\\
&&\nb \left.+2\ddot{M}_{12}\cos2\phi\cos\theta-2\ddot{M}_{13}\cos{\phi}\sin{\theta}\right.\\
&&\left.+2\ddot{M}_{23}\sin\phi\sin{\theta}\right].
\eqn
Particularly, let us consider a binary system with masses $m_1$ and $m_2$, and also assume that the motion is circular.
For simplification, we also assume that the orbit lies in the $(x,y)$ plane,  such that the relative coordinates $x_0^i$ are
\bqn
x_0&=&R\cos(\omega_st+\frac{\pi}2),\\
y_0&=&R\sin(\omega_st+\frac{\pi}2),\\
z_0&=&0,
\eqn
where $R$, $\omega_s$ are the orbital radius and  frequency,  respectively. Then, we find that $M^{ij}$ is given by $M^{ij}=\mu x_0^ix_0^j$ with $\mu=m_1m_2/(m_1+m_2)$. By using Kepler's law, we finally get that
\bqn
\nb D^{(\gamma)}_{+}&=&\frac{p_c}{{\cal{R}}}\cdot\left(\frac{1+\cos^2\iota}{2}\right)\cdot\cos\left(2\pi f_{gw}t_{ret}+2\phi\right),\nb\\
D^{(\gamma)}_{\times}&=&\frac{p_c}{{\cal{R}}}\cdot\cos\iota\cdot\sin(2\pi f_{gw}t_{ret}+2\phi),
\eqn
with
\bqn
p_c&\equiv&4\left(G_N M_c\right)^{5/3}\left(\pi f_{gw}\right)^{2/3},\nb\\
M_c&=&\mu^{3/5}(m_1+m_2)^{2/5}, \nb\\
 f_{gw}&=&\frac{\omega_{gw}}{2\pi}=\frac{\omega_{s}}{\pi},
\eqn
where $t_{ret}$, $f_{gw}$ and $M_c$ are the retarded time, frequency of GWs and the chirp mass respectively, $\iota$ is the angle between the normal to the orbit and the line-of-light which equals to $\theta$.
On a quasi-circular orbit with  $R=R(t), \; \omega_s=\omega_s(t)$, we define \cite{Maggiore:1900zz,Maggiore:2006uy}
\bqn
\Phi(t)&\equiv&2\int^{t_{ret}}\mathrm{d}t'\omega_s(t')=\int^{t_{ret}}\mathrm{d}t'\omega_{gw}(t'),\nb\\
 &=&-2\left(5G_NM_c\right)^{-5/8}\tau^{5/8}+\Phi_0,
\eqn
where $\tau$ is the time interval to binary's coalescence with $\tau\equiv(t_{c})_{ret}-t_{ret}=t_{c}-t$ , and $\Phi_0$ denotes an integration constant. Defining a corresponding time interval $\hat{\tau}\equiv(\eta_{c})_{ret}-\eta_{ret}=\sqrt{2}(v_c-v)$,
  \protect\footnote{  $\eta_{ret}\equiv\eta-\left|\vec{x}\right|=\sqrt{2}v$ is the retarded time, $\eta_c$ is the value of $\eta$ at the coalescence while $v_c$ is equal to the corresponding retarded time, and $\hat{\tau}$ denotes the time interval to the coalescence.}
    we find that
\bqn
\nb
D^{(\gamma)}_{+}(\hat{\tau})&=&\frac{\hat{p}_c(\hat{\tau})}{{\cal{R}}}\cdot\Big(\frac{1+\cos^2\iota}{2}\Big)\cdot\cos\left[\Phi(\hat{\tau})\right],\nb\\
D^{(\gamma)}_{\times}(\hat{\tau})&=&\frac{\hat{p}_c(\hat{\tau})}{{\cal{R}}}\cdot\cos\iota\cdot\sin\left[\Phi(\hat{\tau})\right],
\eqn
with
\bqn
\nb
\Phi(\hat{\tau})&\equiv& -2\left(5G_NM_c\right)^{-5/8}\left(\frac{\hat{\tau}}{1+z_s}\right)^{5/8}+\Phi_0,\nb\\
\hat{p}_c(\hat{\tau})&\equiv&4\left(G_N M_c\right)^{5/3}\left(\pi f_{gw}(\hat{\tau})\right)^{2/3},\nb\\
f_{gw}(\hat{\tau})&=&\frac1{\pi}\left(\frac{5}{256}\frac{1+z_s}{\hat{\tau}}\right)^{3/8}\left(G_NM_c\right)^{-5/8},
\eqn
where $\hat{\tau}\equiv\tau (1+z_s),\;  d\eta=dt/a(t)=(1+z_s)dt$, with $z_s$ being the redshift of the source.
From Eq.(\ref{2.2}), it can be shown that
\bqn
\lb{P}
\nb {\cal{P}}_{+} &=&\hat{p}_c(\hat{\tau})\cdot\left(\frac{1+\cos^2\iota}{2}\right)\cdot\cos\left[\Phi(\hat{\tau})\right],\nb\\
{\cal{P}}_{\times} &=& \hat{p}_c(\hat{\tau})\cdot\cos\iota\cdot\sin\left[\Phi(\hat{\tau})\right].
\eqn
By a direct calculation, Eq.(\ref{gv}) can be obtained. During these epochs, we assume that $z \gg x, y$ and $\iota=0$, so that $\left|\vec{x}\right| \simeq |z|$, where $x$ and $y$ are of the order of the size of the source as shown in
Fig. \ref{fig:Pds}.

\end{document}